\documentclass[journal]{IEEEtran}

\usepackage{cite}
\usepackage{amsmath,amssymb,amsfonts}
\usepackage{algorithmic}
\usepackage{graphicx}
\usepackage{textcomp}
\usepackage{xcolor}
\usepackage{subcaption, bm}
\usepackage{stfloats}
\usepackage[colorlinks=true,
            linkcolor=blue,
            citecolor=blue,
            urlcolor=blue]{hyperref}
\usepackage{algorithm}
\usepackage{booktabs}
\usepackage{multirow}
\usepackage{makecell}

\newtheorem{lemma}{Lemma}
\newtheorem{theorem}{Theorem}
\newtheorem{remark}{Remark}

\begin{document}

\bstctlcite{BSTcontrol}

\title{AFDMA with Heterogeneous Post-Chirp Parameters: Supporting User-Specific Waveform Configurations}

\author{Hengyou Kong,
        Xiaoli Xu,~\IEEEmembership{Member,~IEEE,}
        Zilong Liu,~\IEEEmembership{Senior Member,~IEEE,}
        Yanqun Tang,~\IEEEmembership{Member,~IEEE,}
        and~Yong Zeng,~\IEEEmembership{Fellow,~IEEE}
\thanks{H. Kong, X. Xu and Y. Zeng are with the National Mobile Communication Research Laboratory, Southeast University, Nanjing 210096, China. Y. Zeng is also with the Purple Mountain Laboratories, Nanjing 211111, China (e-mail: $\{$230268151, xiaolixu, yong$\_$zeng$\}$@seu.edu.cn). \textit{(Corresponding author: Yong Zeng.)}}
\thanks{Z. Liu is with the School of Computer Science and Electronics Engineering, University of Essex, CO4 3SQ Colchester, U.K. (e-mail: zilong.liu@essex.ac.uk).}
\thanks{Y. Tang is with the School of Electronics and Communication Engineering, Sun Yat-sen University, Shenzhen 518107, China. (e-mail: tangyq8@mail.sysu.edu.cn).}
}

\IEEEaftertitletext{\vspace{-2.5em}}
\maketitle

\begin{abstract}
This paper investigates a downlink affine frequency division multiple access (AFDMA) framework supporting flexible user-specific waveform configurations with heterogeneous post-chirp parameters. The proposed framework is in sharp contrast to conventional disjoint chirp-subcarrier allocation, which may destroy inter-user orthogonality when users employ heterogeneous post-chirp parameters. Such a loss of orthogonality manifests as cross-basis leakage onto user-allocated chirp subcarriers, which we characterize by analyzing the cross-projection between user-specific discrete affine Fourier transform bases. Specifically, we express the cross-projection entries as quadratic Gauss sums whose quadratic coefficients depend on pairwise post-chirp differences. This yields a sufficient pairwise residue-class orthogonality rule that preserves inter-user orthogonality by satisfying the leakage-free condition. The resulting chirp-subcarrier allocation problem with pairwise modular constraints is reduced to the residue-class level. To solve this problem, we develop a two-stage framework that recursively searches for a feasible representative allocation and deterministically assigns the remaining residue classes. We prove that the problem is either infeasible or admits an optimal full-loading solution. Simulations show that the proposed allocation avoids the bit error rate (BER) floors and signal-to-interference-plus-noise ratio (SINR) ceiling of conventional localized allocation, achieves near-single-user BER, and provides comparable BER while reducing receiver complexity compared with conventional common-maximum-post-chirp AFDMA.
\end{abstract}


%
\IEEEpeerreviewmaketitle

\section{Introduction}

\IEEEPARstart{N}{ext}-generation networks are expected to serve a wide range of wireless terminals that differ in quality-of-service requirements, deployment environments, and device capabilities \cite{zeng2024tutorial,chen2025integrated,zhou2026scenario}. A representative integrated aerial--terrestrial network may simultaneously support low-mobility ground devices and high-mobility aerial terminals such as unmanned aerial vehicles (UAVs), airplanes, and low-Earth-orbit (LEO) satellites \cite{itu2023imt2030}. Accommodating such heterogeneity calls for highly flexible waveform configurations and user-specific parameter adaptation according to individual channel conditions, transceiver capabilities, spectral constraints, and communication/sensing requirements \cite{bir2025orthogonal}.

Against the aforementioned background, not every waveform is suitable for such heterogeneous operating conditions. Orthogonal frequency division multiplexing (OFDM) offers mature implementation and low-complexity equalization, but Doppler spread may severely degrade subcarrier orthogonality and increase the error rate in high-mobility doubly selective channels \cite{wang2026flexible}. Orthogonal chirp division multiplexing (OCDM) may outperform uncoded OFDM, but it does not always attain full diversity over arbitrary doubly selective channels \cite{bemani2023affine}. Orthogonal time frequency space (OTFS) exploits channel diversity in the delay-Doppler (DD) domain \cite{hadani2017orthogonal}, yet it incurs considerable processing complexity and high pilot overhead \cite{zhang2026unified}. Delay-Doppler alignment modulation (DDAM) reduces the channel DD spread through DD compensation and path-based beamforming \cite{lu2023delay,xiao2024integrated,liu2024orthogonal}, but it requires accurate path-specific channel state information (CSI) \cite{xiao2024rethinking}, which is challenging in practical implementations.

Among many others, affine frequency division multiplexing (AFDM) \cite{bemani2023affine} is attractive for coping with heterogeneous services, devices, and wireless environments due to its parameterized waveform structure \cite{li2024chirp,sui2026mimo,yin2025affine}. AFDM employs multiple orthogonal chirp subcarriers whose basis functions are flexibly configured by the pre- and post-chirp parameters $c_2$ and $c_1$, respectively \cite{rou2026affine}. Building upon the discrete affine Fourier transform (DAFT), it also subsumes OFDM and OCDM as special cases, highlighting its flexibility \cite{bemani2023affine}. 
Based on AFDM, affine frequency division multiple access (AFDMA) supports multiple users by allocating their data symbols over different chirp subcarriers \cite{tao2024daft,tao2025affine,yin2025downlink}. Most existing AFDMA designs employ a common $c_1$ for all users. With a common $c_1$, mutually disjoint chirp-subcarrier sets preserve transmitter-side inter-user orthogonality, thereby permitting conventional resource-allocation schemes such as localized allocation with a contiguous chirp-subcarrier block for each user \cite{tao2024daft,tao2025affine,yin2022pilot}. However, enforcing a common $c_1$ prevents users from selecting their post-chirp parameters according to user-specific considerations, which include:

\emph{1) Channel adaptation:} The post-chirp parameter controls DD path separation in the DAFT domain. A sufficiently large $c_1$ can prevent equivalent channel paths from overlapping, thus enabling AFDM to achieve full channel diversity. Generally, high-mobility users may require larger $c_1$ values \cite{bemani2023affine}.

\emph{2) Receiver complexity and spectral constraints:} A larger post-chirp parameter can increase the span of significant entries in the DAFT-domain channel matrix \cite{bemani2022low} and hence increase the receiver complexity, e.g., in the sparsity-exploiting low-complexity banded minimum mean square error (LC-BMMSE) receiver \cite{mao2026joint}. It can also aggravate spectral spreading and out-of-band (OOB) emissions \cite{mirabella2026continuous}. Processing-constrained users may therefore favor smaller $c_1$ values.

\emph{3) Sensing requirements:} The post-chirp parameter shapes the ambiguity-function geometry and its associated unambiguous region. Sensing-enabled users may favor different $c_1$ values according to their sensing objectives \cite{yin2025ambiguity,kong2026impact}.

Consequently, no single post-chirp parameter is universally preferable in a multi-user system. In an integrated aerial–terrestrial network, for example, high-mobility aerial users, receiver-complexity-constrained users, and sensing-enabled users may prefer different $c_1$ values. Furthermore, for compatibility with existing networks, coexistence of AFDM and OFDM/OCDM users also raises the issue of different post-chirp parameters. User-specific post-chirp parameter adaptation is therefore desirable.

However, heterogeneous-$c_1$ transmission can pose a fundamental challenge to inter-user orthogonality. With heterogeneous $c_1$ values, a chirp subcarrier in one user-specific DAFT basis generally projects onto multiple subcarriers in another. Such cross-basis leakage may overlap user-allocated chirp subcarriers, so conventional disjoint allocation no longer guarantees inter-user orthogonality and may produce inter-user interference (IUI) even before accounting for channel effects.

Recent studies have moved beyond the common-$c_1$ assumption. Orthogonal AFDM-OFDM coexistence is achieved using specifically matched AFDM parameters and residue-class allocation \cite{bir2025orthogonal}. AFDM-enabled rate-splitting multiple access was studied in \cite{li2026chirp} by optimizing user-specific chirp parameters to improve the cooperative diversity gain and suppress multi-user interference. But this design is based on a non-orthogonal transmission architecture with cooperative detection. To the best of our knowledge, no research work has provided a general orthogonal resource-allocation rule for AFDMA users with prescribed heterogeneous $c_1$ values. To fill such a gap, this paper develops a downlink AFDMA framework supporting user-specific waveform configurations with heterogeneous post-chirp parameters. The key idea is to preserve inter-user orthogonality through residue-class-based chirp-subcarrier allocation based on cross-projection matrices characterized by quadratic Gauss sums. The main contributions are summarized as follows.

\begin{itemize}
    \item We establish a downlink AFDMA transceiver framework supporting user-specific post-chirp parameters within the same AFDM block. The framework accommodates post-chirp parameters selected according to individual channel conditions, receiver complexity constraints, or sensing requirements while preserving inter-user orthogonality.
    \item We characterize the cross-basis leakage induced by heterogeneous post-chirp parameters through the cross-projection between user-specific DAFT bases, providing insight into orthogonal chirp-subcarrier allocation. The cross-projection entries are expressed in terms of quadratic Gauss sums whose quadratic coefficients are determined by the pairwise post-chirp differences. These sums link cross-basis leakage to congruence relations between chirp-subcarrier indices, yielding a sufficient residue-class orthogonality rule with a user-pair-dependent effective modulus that preserves inter-user orthogonality. We further analyze the average cross-basis leakage energy and derive the symbol-wise output SINR under practical channel equalization.
    \item We formulate chirp-subcarrier allocation for heterogeneous post-chirp AFDMA as an optimization problem with pairwise modular constraints and reduce it to the residue-class level. We solve the reduced problem using a two-stage allocation framework comprising a recursive search for a feasible representative allocation and a deterministic extension that assigns the remaining fine residue classes. It is proved that any feasible representative allocation can be extended to an optimal full-loading allocation, establishing the infeasibility-or-optimal-full-loading property.
\end{itemize}

Numerical results show that the proposed allocation avoids the BER floors and SINR ceiling of conventional localized allocation and achieves near-single-user performance. Compared with conventional common-maximum-$c_1$ AFDMA, it provides comparable BER with lower receiver complexity. The robustness of the proposed framework to imperfect CSI is also evaluated.

The remainder of this paper is organized as follows. Section II presents the downlink AFDMA system model with heterogeneous post-chirp parameters. Section III analyzes the cross-basis leakage pattern and derives the residue-class orthogonality condition. Section IV develops the two-stage residue-class allocation framework. Section V presents the simulation results, and Section VI concludes the paper.

\textit{Notation}: $(\cdot)^T$, $(\cdot)^H$, $(\cdot)^*$, $\left\|\cdot\right\|$, $\left\lceil\cdot\right\rceil$, and $\odot$ denote the transpose, the Hermitian transpose, the conjugate, the Euclidean norm, the ceiling operation, and the Hadamard product, respectively. $\mathbb{Z}$ and $\mathbb{N}$ denote the sets of integers and positive integers, respectively. $\mathbf{I}_N$ and $\mathbf{0}_N$ represent the $N\times N$ identity matrix and the $N\times 1$ all-zero vector, respectively. $\mathrm{diag}(\cdot)$ transforms a vector into a diagonal matrix. $\mathrm{tr}(\cdot)$ denotes the trace. $a\mid b$ and $a\nmid b$ denote that $a$ divides and does not divide $b$, respectively. $a\equiv b\pmod N$ and $a\not\equiv b\pmod N$ denote that $a$ and $b$ are and are not congruent modulo $N$, respectively. $|\mathcal{A}|$, $\mathcal{A}\setminus\mathcal{B}$, and $\bigcup_{u=1}^{U}\mathcal{A}_u$ denote the cardinality of set $\mathcal{A}$, the set difference between $\mathcal{A}$ and $\mathcal{B}$, and the union of $\{\mathcal{A}_u\}_{u=1}^{U}$, respectively. All other notation is defined as needed.



\section{System Model}
\label{Section_II}

\begin{figure}[!t]
    \centering
    \includegraphics[width=0.70\linewidth]{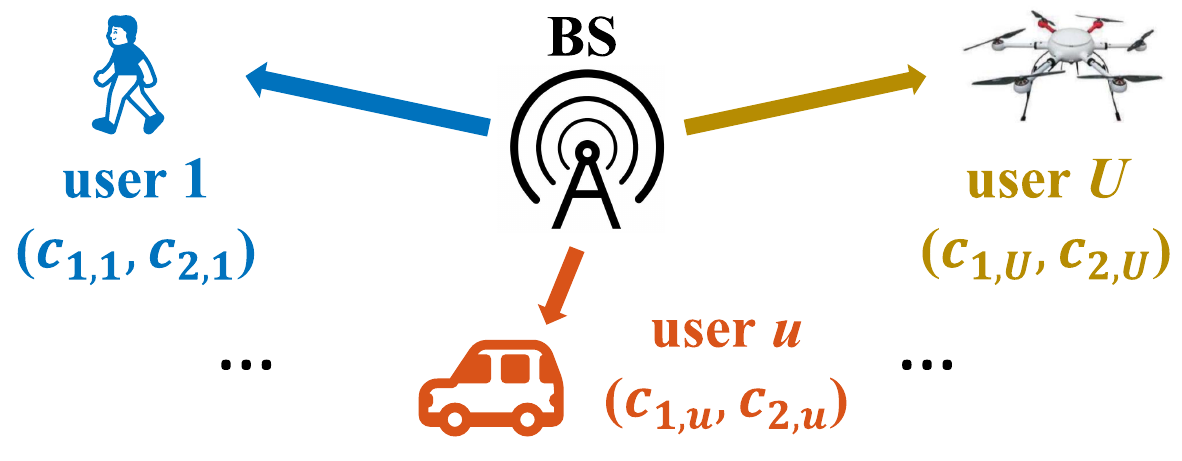}
    \caption{Downlink AFDMA transmission with heterogeneous post-chirp parameters.}
    \label{fig:downlink_heterogeneous_c1_transmission}
\end{figure}

\begin{figure}[!t]
    \centering
    \includegraphics[width=1.00\linewidth]{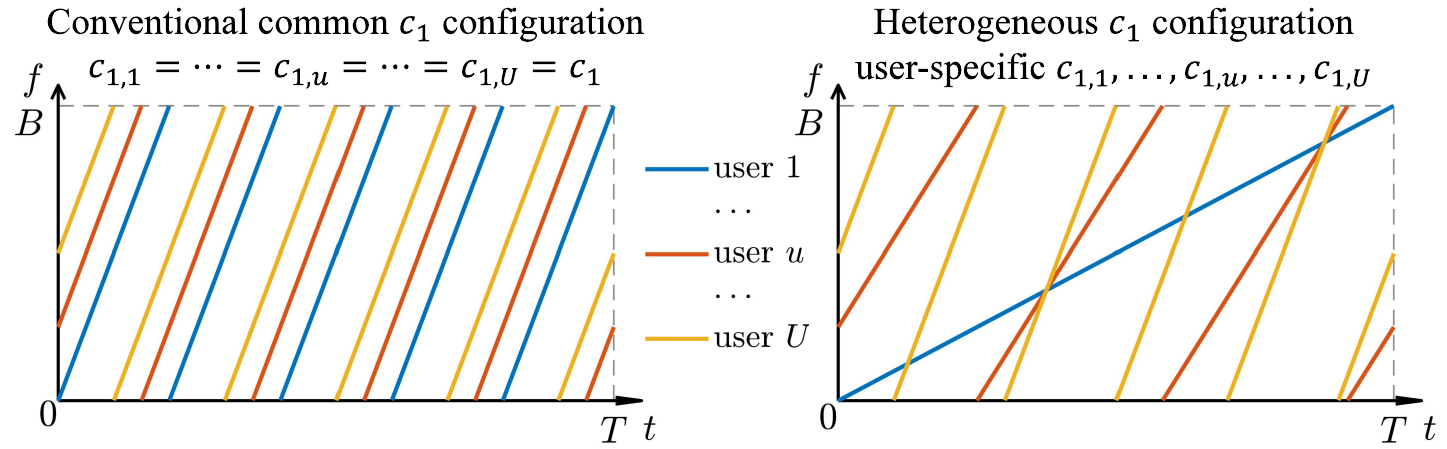}
    \caption{Time-frequency representations of user-assigned chirp subcarriers under common and heterogeneous post-chirp configurations.}
    \label{fig:TF_domain}
\end{figure}

Fig.~\ref{fig:downlink_heterogeneous_c1_transmission} illustrates an integrated aerial-terrestrial downlink scenario in which a base station (BS) serves $U$ users with AFDMA. User $u$ employs post- and pre-chirp parameters $c_{1,u}$ and $c_{2,u}$, respectively. As shown in Fig.~\ref{fig:TF_domain}, the conventional configuration imposes a common $c_1$ on all users, resulting in parallel chirp-subcarrier representations with the same time-frequency slope. However, this restriction precludes user-specific adaptation of $c_1$ to individual channel, implementation, or sensing requirements. In the heterogeneous configuration, the prescribed $c_{1,u}$ values can instead vary across users according to user-specific requirements, producing different slopes and intersecting representations. The resulting mismatch among the user-specific DAFT bases challenges inter-user orthogonality. This work focuses on enabling the orthogonal coexistence of these user-specific waveform configurations. The corresponding transceiver structures and signal processing procedures are illustrated in the following subsections.

\subsection{Downlink Transmitter}

\begin{figure*}[!htbp]
    \centering
    \includegraphics[width=1\linewidth]{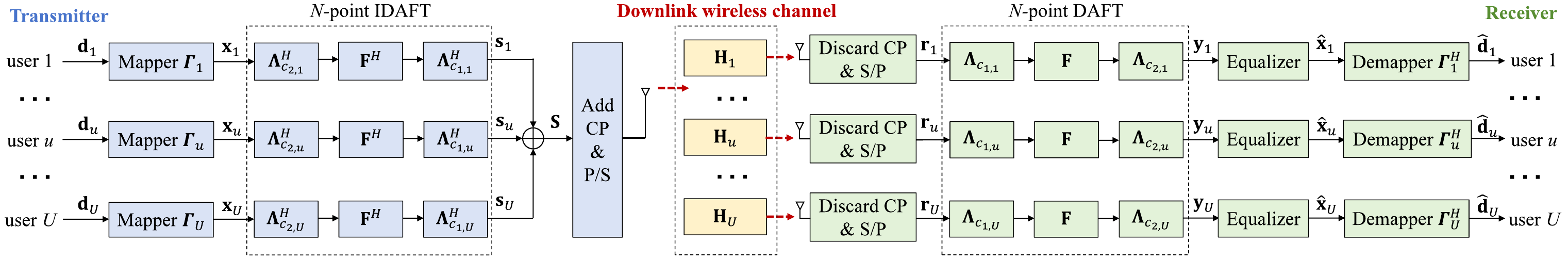}
    \caption{Downlink AFDMA transceiver structure with heterogeneous post-chirp parameters.}
    \label{fig:downlink_heterogeneous_c1_transceiver}
\end{figure*}

The structure of the downlink heterogeneous-$c_1$ AFDMA transmitter is shown in Fig.~\ref{fig:downlink_heterogeneous_c1_transceiver}. Let $\mathbf{d}_u=[d_{0,u},\ldots,d_{N_u-1,u}]^T\in\mathbb{C}^{N_u\times 1}$ denote the data-symbol vector of user $u$, where $N_u$ is the number of chirp subcarriers allocated to this user. The modulation symbols $d_{n,u}$ are assumed to be independent and identically distributed.

Next, $\mathbf{d}_u$ is mapped to the allocated DAFT-domain chirp subcarriers of user $u$. Let $\mathbb{Z}_N=\{0,1,\ldots,N-1\}$ be the index set of the $N$ chirp subcarriers and $\mathcal{S}_u=\{\mu_{0,u},\ldots,\mu_{N_u-1,u}\}\subseteq\mathbb{Z}_N$ be the chirp subcarrier index set allocated to user $u$, where $|\mathcal{S}_u|=N_u$.
Let $\mathbf{e}_n$ denote the $n$-th column of $\mathbf{I}_N$.
If $d_{i,u}$ is mapped to the $\mu_{i,u}$-th chirp subcarrier, the mapping matrix of user $u$ is defined as
\begin{equation}
    \mathbf{\Gamma}_u
    =
    [\mathbf{e}_{\mu_{0,u}},\ldots,\mathbf{e}_{\mu_{N_u-1,u}}]
    \in\mathbb{C}^{N\times N_u},
\end{equation}
and the demapping matrix $\mathbf{\Gamma}_u^H$ satisfies $\mathbf{\Gamma}_u^H\mathbf{\Gamma}_u=\mathbf{I}_{N_u}$.
The resulting DAFT-domain transmit vector $\mathbf{x}_u\in\mathbb{C}^{N\times 1}$ is
\begin{equation}\label{equ:x_u}
    \mathbf{x}_u
    =
    \sum_{i=0}^{N_u-1} d_{i,u}\mathbf{e}_{\mu_{i,u}}
    =
    \mathbf{\Gamma}_u\mathbf{d}_u.
\end{equation}

The chirp subcarrier mapping is followed by the $N$-point inverse DAFT (IDAFT) to generate the time-domain signal $\mathbf{s}_u$, given by
\begin{equation}
    \mathbf{s}_u = \mathbf{\Lambda}_{c_{1,u}}^H\mathbf{F}^H\mathbf{\Lambda}_{c_{2,u}}^H\mathbf{x}_u,
\end{equation}
where $\mathbf{F}$ is the normalized discrete Fourier transform (DFT) matrix and $\mathbf{\Lambda}_{c}$ is a diagonal chirp matrix with $\mathbf{\Lambda}_{c}=\mathrm{diag}(e^{-\imath2\pi cn^2},n=0,\dots,N-1)$. Define the user-specific DAFT matrix as
\begin{equation}\label{equ:A_u}
\mathbf{A}_u=\mathbf{\Lambda}_{c_{2,u}}\mathbf{F}\mathbf{\Lambda}_{c_{1,u}}.
\end{equation}
Since $\mathbf A_u$ is unitary, its rows, or equivalently the columns of $\mathbf A_u^H$, form an orthonormal user-specific DAFT basis \cite{bemani2023affine}. Hence, different chirp parameters generally yield different DAFT bases. While user-specific $c_{2,u}$ values have been considered in multiple-access AFDM transmission \cite{zheng2025partially}, this work allows downlink AFDMA users to employ different post-chirp parameters $c_{1,u}$. Following the conventional discrete AFDM design \cite{bemani2023affine}, we set $c_{1,u}={a_{u}}/{(2N)},\, a_u\in\mathbb{Z}$ for each user.

The BS then forms the downlink AFDMA transmit block $\mathbf{S}$ by superimposing the user-specific time-domain signals, which can be expressed as
\begin{equation}\label{equ:downlink_transmit_block}
    \mathbf{S}=\sum_{u=1}^{U}\mathbf{s}_u.
\end{equation}

Finally, the cyclic prefix (CP)\footnote{The chirp-periodic prefix (CPP) used in AFDM reduces to the conventional CP when $2Nc_{1,u}$ is an integer and $N$ is even. Both conditions hold in this work, so the CP appended to $\mathbf{S}$ is equivalent to the CPP for each user.} longer than the maximum delay spread among all users is appended to $\mathbf{S}$ to prevent inter-block interference. After parallel-to-serial conversion, the BS broadcasts the resulting AFDMA block $\mathbf{S}$ to all users.

\subsection{Channel Model and User-Specific Downlink Receiver}
After serial-to-parallel conversion and CP removal, the time-domain signal received by user $u$ is 
\begin{equation}
    \mathbf{r}_u = \mathbf{H}_u\mathbf{S}+\mathbf{w}_u=\mathbf{H}_u\sum_{j=1}^{U}\mathbf{A}_j^H\mathbf{x}_j+\mathbf{w}_u,
\end{equation}
where $\mathbf{H}_u\in\mathbb{C}^{N\times N}$ is the downlink channel matrix of user $u$, and $\mathbf{w}_u\sim\mathcal{CN}\left(\mathbf{0}_N,\sigma_u^2\mathbf{I}_N\right)$ is an additive white Gaussian noise (AWGN) vector.

We model the channel of user $u$ as a general integer-delay and fractional-Doppler channel with $P_u$ propagation paths. The $i$-th path of user $u$ has a complex channel gain $h_{i,u}$, a delay $\tau_{i,u}$ and a Doppler shift $\nu_{i,u}$. Given the sampling period $T_s$ and subcarrier spacing $\Delta f={1}/{(NT_s)}$, the normalized delay and Doppler shift of the $i$-th path are defined as 
\begin{equation}
l_{i,u}=\frac{\tau_{i,u}}{T_s},\qquad
\alpha_{i,u}=\frac{\nu_{i,u}}{\Delta f}=k_{i,u}+\kappa_{i,u},    
\end{equation}
where $l_{i,u}\in[0,l_{u,\max}]$ and $k_{i,u}\in[-k_{u,\max},k_{u,\max}]$ are the integer components, while $\kappa_{i,u}\in[-0.5,0.5)$ is the fractional Doppler component. The corresponding channel matrix is
\begin{equation}\label{equ:channel_model}
    \mathbf{H}_u
    =
    \sum_{i=1}^{P_u}h_{i,u}\mathbf{\Delta}_{\alpha_{i,u}}\mathbf{\Pi}^{l_{i,u}},
\end{equation}
where $\mathbf{\Delta}_{\alpha_{i,u}}=\mathrm{diag}(e^{-\imath2\pi \alpha_{i,u}n/N},n=0,\dots,N-1)$ is the Doppler phase-rotation matrix, and $\mathbf{\Pi}$ is the forward cyclic-shift matrix.

At the receiver, user $u$ demodulates $\mathbf{r}_u$ with its own DAFT matrix $\mathbf{A}_u$. Let $\bar{\mathbf{H}}_u=\mathbf{A}_u\mathbf{H}_u\mathbf{A}_u^H$ denote the effective channel matrix. The received DAFT-domain signal is then
\begin{equation}\label{equ:yrec_sig}
\small
        \mathbf{y}_u
        =
        \mathbf{A}_u\mathbf{r}_u
        =
        \underbrace{
        \bar{\mathbf{H}}_u\mathbf{x}_u
        }_{\text{desired signal}}
        +
        \underbrace{
        \bar{\mathbf{H}}_u
        \sum\nolimits_{j\neq u}
        \mathbf{A}_u\mathbf{A}_j^H\mathbf{x}_j
        }_{\text{IUI}} 
        +
        \tilde{\mathbf{w}}_u,
\end{equation}
where $\tilde{\mathbf{w}}_u=\mathbf{A}_u\mathbf{w}_u$ is the DAFT-domain noise vector. Since $\mathbf{A}_u$ is unitary, $\tilde{\mathbf{w}}_u$ has the same statistical properties as ${\mathbf{w}}_u$.

The first two terms in \eqref{equ:yrec_sig} are the desired signal and the IUI after user-specific DAFT demodulation, respectively. When all users employ identical chirp parameters, their signals share a common DAFT basis and the cross-projection term $\mathbf{A}_u\mathbf{A}_j^H=\mathbf{I}_N$. Thus, the cross-projection introduces no additional effect, and the IUI is shaped by the downlink channel. With heterogeneous chirp parameters, $\mathbf{A}_u\mathbf{A}_j^H$ generally differs from $\mathbf{I}_N$. Consequently, both the cross-projection between user-specific DAFT matrices and the downlink channel affect the IUI.

\section{Cross-Basis Leakage Pattern and Residue-Class Orthogonality Analysis}
\label{Section_III}

As shown in Section~\ref{Section_II}, the received signal of user $u$ contains the desired signal and the IUI caused by the superposition of heterogeneous-$c_1$ AFDM signals. This section characterizes the cross-basis leakage underlying this IUI and derives a sufficient residue-class orthogonality rule that preserves inter-user orthogonality in downlink heterogeneous-$c_1$ AFDMA.
\subsection{Cross-Basis Leakage Pattern}
\label{Subsection_III_A}

To simplify the notation in \eqref{equ:yrec_sig}, define
\begin{equation}\label{equ:p_u}
    \mathbf{p}_u = \sum_{j\neq u}\mathbf{A}_u\mathbf{A}_j^H\mathbf{x}_j=\sum_{j\neq u}\mathbf{T}_{u,j}\mathbf{x}_j,
\end{equation}
where $\mathbf{T}_{u,j}=\mathbf{A}_u\mathbf{A}_j^H$ is the cross-projection matrix from the DAFT domain of user $j$ to that of user $u$. The vector $\mathbf p_u$ denotes the cross-basis leakage observed in the DAFT domain of user $u$. It is induced by the mismatch between the user-specific DAFT bases and is independent of the channel parameters. Using \eqref{equ:p_u}, the received signal in \eqref{equ:yrec_sig} becomes
\begin{equation}\label{equ:yrec_sig_II}
    \mathbf{y}_u
    =
    \bar{\mathbf{H}}_u
    (\mathbf{x}_u+\mathbf{p}_u)
    +
    \tilde{\mathbf{w}}_u.
\end{equation}

To isolate the cross-basis leakage pattern from channel effects, we consider ideal zero-forcing (ZF) equalization $\mathbf W_u$ satisfying $\mathbf W_u\bar{\mathbf H}_u=\mathbf I_N$ and temporarily neglect the equalized noise. The equalized DAFT-domain signal is then given by
\begin{equation}\label{equ:equlaized}
    \hat{\mathbf{x}}_u
    =
    \mathbf{W}_u\mathbf{y}_u
    =
    \mathbf{x}_u+\mathbf{p}_u.
\end{equation}
Chirp-subcarrier demapping then yields
\begin{equation}\label{equ:demapped}
    \hat{\mathbf{d}}_u
    =
    \mathbf{\Gamma}_u^H\hat{\mathbf{x}}_u
    =
    \mathbf{d}_u+\mathbf{\Gamma}_u^H\mathbf{p}_u.
\end{equation}
By \eqref{equ:demapped}, IUI remains even after ideal ZF equalization whenever $\mathbf{\Gamma}_u^H\mathbf p_u\neq\mathbf 0_{N_u}$, indicating that inter-user orthogonality is not preserved. 
To preserve such orthogonality, the allocation must prevent cross-basis leakage on the chirp subcarriers allocated to user $u$.
For nonzero modulation symbols, this requirement can be equivalently expressed as
\begin{equation}\label{equ:x_hadamard_p}
    \mathbf{\Gamma}_u^H\mathbf{p}_u=\mathbf{0}_{N_u}
    \Longleftrightarrow
    \mathbf{x}_u\odot\mathbf{p}_u=\mathbf{0}_N,\quad\forall u.
\end{equation}
We refer to this requirement as the leakage-free condition.

\subsection{Chirp-Subcarrier-Wise Orthogonality Condition}
Let $x_{r,u}$ and $p_{r,u}$ denote the $r$-th entries of $\mathbf x_u$ and $\mathbf p_u$, respectively. According to \eqref{equ:x_u}, for nonzero modulation symbols, $x_{r,u}\neq0$ if and only if $r\in\mathcal{S}_u$. Thus, \eqref{equ:x_hadamard_p} is equivalent to 
\begin{equation}
    p_{r,u}=0,\quad \forall r\in\mathcal{S}_u,\ \forall u.
\end{equation}

From \eqref{equ:p_u}, the $r$-th entry of $\mathbf{p}_u$ is given by
\begin{equation}
    p_{r,u}
    =
    \sum_{j\neq u}
    \sum_{q\in\mathcal{S}_j}
    \mathbf{T}_{u,j}[r,q]x_{q,j}.
\end{equation}
Guaranteeing $p_{r,u}=0$ for arbitrary transmitted data symbols requires that
\begin{equation}\label{equ:cross-projection}
    \mathbf{T}_{u,j}[r,q]=0,
    \quad
    \forall r\in\mathcal{S}_u,\ 
    \forall q\in\mathcal{S}_j,\ 
    \forall u\neq j.
\end{equation}
Here, $\mathbf{T}_{u,j}[r,q]$ represents the cross-projection from the $q$-th chirp subcarrier allocated to user $j$ onto the $r$-th chirp subcarrier allocated to user $u$. The leakage-free condition ensuring inter-user orthogonality in \eqref{equ:x_hadamard_p} is equivalent to the chirp-subcarrier-wise orthogonality requirement in \eqref{equ:cross-projection}.

From the definition of $\mathbf{A}_u$ in \eqref{equ:A_u}, the $(r,q)$-th element of $\mathbf{T}_{u,j}$ is derived as
\begin{equation}\label{equ:T_u_j}
    \begin{aligned}
        \mathbf{T}_{u,j}[r,q]
        &= \frac{1}{N}
        e^{\imath 2\pi (c_{2,j}q^2-c_{2,u}r^2)}    \\
        &\quad \times
        \sum_{n=0}^{N-1}
            e^{\imath 2\pi (c_{1,j}-c_{1,u})n^2}  
            e^{\imath 2\pi (q-r)n/N}.
    \end{aligned}
\end{equation}
Equation \eqref{equ:T_u_j} reveals two important properties of the chirp-subcarrier-wise cross-projection between the user-specific DAFT domains.

\begin{itemize}
    \item{The user-specific $c_1$ values determine the summation term and hence the magnitude and zero/nonzero pattern of $\mathbf T_{u,j}[r,q]$. By contrast, the user-specific $c_{2}$ values contribute only the phase factor $e^{\imath 2\pi (c_{2,j}q^2-c_{2,u}r^2)}$}.
    \item{Mathematically, with $c_{1,j}=c_{1,u}$, \eqref{equ:T_u_j} reduces to
    \begin{equation}
        \begin{aligned}
            \mathbf{T}_{u,j}[r,q]=
            e^{\imath2\pi(c_{2,j}q^2-c_{2,u}r^2)}\delta_N(q-r),
        \end{aligned}
    \end{equation}
    where $\delta_N(\cdot)$ denotes the modulo-$N$ Kronecker delta. When $c_{1,j}\neq c_{1,u}$, the summation term becomes a quadratic Gauss sum whose zero pattern is different.}
\end{itemize}

Hence, for common-$c_1$ AFDMA, the Kronecker delta structure makes disjoint DAFT-domain chirp-subcarrier allocation sufficient for inter-user orthogonality.
By contrast, in heterogeneous-$c_1$ AFDMA, the cross-projection is governed by a quadratic Gauss sum, so conventional disjoint chirp-subcarrier allocation no longer guarantees inter-user orthogonality and a new resource-allocation principle is required.

\subsection{Quadratic-Gauss-Sum-Based Residue-Class Orthogonality}
As derived in \eqref{equ:T_u_j}, the zero/nonzero pattern of $\mathbf{T}_{u,j}[r,q]$ is determined by the quadratic Gauss sum term. According to the conventional AFDM design, $c_{1,u}={a_{u}}/{(2N)},\, a_u\in\mathbb{Z}$. Let 
\begin{equation}
    K_{u,j} = 2N(c_{1,u}-c_{1,j}) \in\mathbb{Z}.    
\end{equation}
The quadratic Gauss sum term in \eqref{equ:T_u_j} is then expressed as
\begin{equation}\label{equ:G_u_j}
    \begin{aligned}
        \mathcal{G}_{u,j}[r,q]
        =
        \sum_{n=0}^{N-1}
        e^{-\imath\frac{2\pi}{N}
        \left(
        (r-q) n + \frac{K_{u,j}}{2}n^2
        \right)}.
    \end{aligned}
\end{equation}

Define 
\begin{equation}\label{equ:D_u_j}
    D_{u,j}=\mathrm{gcd}(|K_{u,j}|,N),    
\end{equation}
where $\mathrm{gcd}(a,b)$ denotes the greatest common divisor of $a$ and $b$. The following lemma characterizes the zero/nonzero pattern of $\mathcal{G}_{u,j}[r,q]$.

\begin{lemma}
\label{lemma1}
    Suppose that the number of chirp subcarriers satisfies $N=2^n,\,n\in\mathbb{N}$. If $D_{u,j} < N$, then
    \begin{equation}\label{equ:lemma1}
        \left|\mathcal{G}_{u,j}[r,q]\right|^2
        =
        \begin{cases}
            N D_{u,j}, & r\equiv q \pmod{D_{u,j}},\\
            0, & r\not\equiv q \pmod{D_{u,j}},
        \end{cases}.
    \end{equation}
\end{lemma}

\textit{Proof:} See Appendix~\ref{app:lemma1}. \hfill$\blacksquare$

\textit{Lemma \ref{lemma1}} provides insight into the chirp-subcarrier-wise orthogonality of heterogeneous-$c_1$ downlink AFDMA. Since $\mathbf{T}_{u,j}[r,q]$ and $\mathcal{G}_{u,j}[r,q]$ have the same zero/nonzero pattern, \textit{Lemma \ref{lemma1}} implies that when $N=2^n$ and $D_{u,j}<N$, $\mathbf{T}_{u,j}[r,q]=0$ if $r\not\equiv q \pmod{D_{u,j}}$. Thus, the $q$-th chirp subcarrier of user $j$ has zero cross-projection onto the chirp subcarriers of user $u$ in different residue classes modulo $D_{u,j}$.

According to \eqref{equ:D_u_j}, one has $D_{u,j}\leq N$. It remains to characterize $D_{u,j}=N$, which is covered by \textit{Lemma \ref{lemma2}}.

\begin{lemma}
\label{lemma2} 
Suppose that the number of chirp subcarriers satisfies $N=2^n,\,n\in\mathbb{N}$, and $D_{u,j}=N$. Then $K_{u,j}/N\in\mathbb{Z}$.

If $K_{u,j}/N$ is even, then
\begin{equation}\label{equ:lemma2_even}
    |\mathcal{G}_{u,j}[r,q]|^2
    =
    \begin{cases}
    N^2, & r\equiv q \pmod{N},\\
    0, & r\not\equiv q \pmod{N}.
    \end{cases}
\end{equation}

If $K_{u,j}/N$ is odd, then
\begin{equation}\label{equ:lemma2_odd}
    |\mathcal{G}_{u,j}[r,q]|^2
    =
    \begin{cases}
        N^2,&r-q\equiv \frac{N}{2} \pmod{N},\\
        0,&r-q\not\equiv \frac{N}{2} \pmod{N}.
    \end{cases}
\end{equation}
\end{lemma}

\textit{Proof:} 
The result follows from Appendix~\ref{app:lemma1} by substituting $D=N$ and $N_0=N/D=1$ into \eqref{equ:appendix_G_A_K_further} and considering whether $K_0=K_{u,j}/N$ is even or odd. \hfill$\blacksquare$

\textit{Lemma \ref{lemma2}} complements \textit{Lemma \ref{lemma1}} by characterizing the remaining case $D_{u,j}=N$, where the chirp-subcarrier-wise orthogonality depends on the parity of $K_{u,j}/N$. 

For even $K_{u,j}/N$, $\mathbf{T}_{u,j}[r,q]=0$ if $r\not\equiv q \pmod N$. This condition is equivalent to $r\neq q$ because $r,q\in\mathbb Z_N$. Hence, the $q$-th chirp subcarrier of user $j$ can have a nonzero cross-projection only onto the same-index chirp subcarrier of user $u$. This case includes the common-$c_1$ setting, where two users only need to occupy different chirp subcarrier indices to avoid cross-basis leakage.

For odd $K_{u,j}/N$, $\mathbf{T}_{u,j}[r,q]=0$ if $r-q\not\equiv N/2 \pmod N$. This means that the $q$-th chirp subcarrier of user $j$ can have a nonzero cross-projection only onto the chirp subcarrier of user $u$ whose index differs by $N/2$ modulo $N$.

Together, \textit{Lemma~\ref{lemma1}} and \textit{Lemma~\ref{lemma2}} show that when $N=2^n$, the pairwise orthogonality is determined by $D_{u,j}$ and, in the special case $D_{u,j}=N$, by the parity of $K_{u,j}/N$. We therefore define the effective pairwise modulus as
\begin{equation}\label{equ:effective_modulus}
    \widetilde{D}_{u,j}=
    \begin{cases}
    D_{u,j}, & D_{u,j}<N,\\
    N, & D_{u,j}=N,\ K_{u,j}/N\ \text{is even},\\
    N/2, & D_{u,j}=N,\ K_{u,j}/N\ \text{is odd}.
    \end{cases}
\end{equation}
The following theorem uses this effective modulus to establish a unified sufficient residue-class orthogonality rule.

\begin{theorem}
\label{theorem1}
Suppose that $N=2^n$. In the downlink heterogeneous-$c_1$ AFDMA system, if the chirp subcarrier sets $\mathcal{S}_u$ and $\mathcal{S}_j$ allocated to users $u$ and $j$ satisfy the pairwise residue-class orthogonality rule
\begin{equation}\label{equ:theorem}
    r\not\equiv q \pmod{\widetilde D_{u,j}},
    \quad
    \forall r\in\mathcal{S}_u,\ 
    \forall q\in\mathcal{S}_j,\
    \forall u< j,
\end{equation}
then the chirp-subcarrier-wise orthogonality requirement in \eqref{equ:cross-projection} holds for all user pairs. Consequently, the transmitter-side leakage-free design principle in \eqref{equ:x_hadamard_p} is satisfied.
\end{theorem}

\textit{Proof:} Since $\mathbf{T}_{u,j}=\mathbf{T}_{j,u}^H$ and $\widetilde{D}_{u,j}=\widetilde{D}_{j,u}$, the orthogonality condition is symmetric for each user pair $(u,j)$. Therefore, it suffices to consider $u<j$.

To show that $\mathbf{T}_{u,j}[r,q]=0$ for all $r\in\mathcal{S}_u$ and $q\in\mathcal{S}_j$, we consider the following three cases:
\begin{itemize}
    \item{If $D_{u,j}<N$, then $\widetilde D_{u,j}=D_{u,j}$. The condition in \eqref{equ:theorem} becomes $r\not\equiv q \pmod{D_{u,j}}$. Hence, according to \textit{Lemma~\ref{lemma1}}, $\mathbf{T}_{u,j}[r,q]=0$.}
    \item{If $D_{u,j}=N$ and $K_{u,j}/N$ is even, then $\widetilde D_{u,j}=N$. The condition in \eqref{equ:theorem} becomes $r\not\equiv q \pmod N$. Hence, according to \textit{Lemma~\ref{lemma2}}, $\mathbf{T}_{u,j}[r,q]=0$.}
    \item{If $D_{u,j}=N$ and $K_{u,j}/N$ is odd, then $\widetilde D_{u,j}=N/2$. The condition in \eqref{equ:theorem} becomes $r\not\equiv q \pmod{N/2}$, i.e., $r-q\not\equiv 0 \pmod{N/2}$. Since $r-q\equiv N/2 \pmod N$ implies $r-q\equiv 0 \pmod{N/2}$, as a contrapositive proposition, $r-q\not\equiv 0 \pmod{N/2}$ is sufficient to ensure the zero case in \eqref{equ:lemma2_odd}, namely $r-q\not\equiv N/2 \pmod N$. Hence, according to \textit{Lemma~\ref{lemma2}}, $\mathbf{T}_{u,j}[r,q]=0$.}
    \hfill$\blacksquare$
\end{itemize}

\textit{Theorem \ref{theorem1}} translates the chirp-subcarrier-wise orthogonality requirement in \eqref{equ:cross-projection} into the sufficient residue-class orthogonality rule in \eqref{equ:theorem}. Under this rule, for each user pair $(u,j)$, disjointness of the residue sets of $\mathcal{S}_u$ and $\mathcal{S}_j$ modulo their effective pairwise modulus $\widetilde D_{u,j}$ is sufficient to guarantee leakage-free transmission and inter-user orthogonality.

\begin{remark}
    The analysis considers the practical fast Fourier transform (FFT)-friendly case $N=2^n$. For arbitrary $N$, the zero/nonzero pattern of \eqref{equ:G_u_j} depends on additional number-theoretic conditions and does not lead to a simpler residue-class allocation rule. Hence, this paper focuses on $N=2^n$.
\end{remark}

\subsection{Average Cross-Basis Leakage Energy and SINR Analysis}
\label{subsection_III_D}

Under the sufficient residue-class orthogonality rule in \textit{Theorem~\ref{theorem1}}, the leakage-free design principle \eqref{equ:x_hadamard_p} holds. $\mathbf{p}_u$ is supported only on the chirp subcarriers not allocated to user $u$, thereby preserving inter-user orthogonality. 

Since $\mathbf{A}_u$ and $\mathbf{A}_j$ are unitary, the
cross-projection matrix $\mathbf{T}_{u,j}=\mathbf{A}_u
\mathbf{A}_j^H$ is also unitary, with
$\mathbf{T}_{u,j}^H\mathbf{T}_{u,j}=\mathbf{I}_N$. According to \eqref{equ:p_u},  the energy of $\mathbf{p}_u$ is given by
\begin{equation}\label{equ:energy_p_u}
    \begin{aligned}
        \left\|\mathbf{p}_u\right\|^2
        =\sum_{j\neq u}\left\|\mathbf{x}_j\right\|^2+\sum_{j\neq u}\sum_{k\neq u,k\neq j}\mathbf{x}_j^H\mathbf{T}_{u,j}^H\mathbf{T}_{u,k}\mathbf{x}_k.
    \end{aligned}
\end{equation}

Since the transmitted signals of different users are statistically independent and have zero mean, i.e.,
$\mathbb{E}\{\mathbf{x}_j\}=\mathbf{0}_N, \forall j$ and
$\mathbb{E}\{\mathbf{x}_k\mathbf{x}_j^H\}=\mathbf{0}_{N\times N}, \forall k\neq j$, the expectation of the cross term in \eqref{equ:energy_p_u} is
\begin{equation}
    \begin{aligned}
        \mathbb{E}\left\{\mathbf{x}_j^H\mathbf{T}_{u,j}^H\mathbf{T}_{u,k}\mathbf{x}_k\right\}=
        \mathrm{tr}\left(
        \mathbf{T}_{u,j}^H
        \mathbf{T}_{u,k}
        \mathbb{E}\{\mathbf{x}_k\mathbf{x}_j^H\}
        \right)
        =0.
    \end{aligned}
\end{equation}
The average energy of the cross-basis leakage vector $\mathbf{p}_u$ is
\begin{equation}\small
    \mathbb{E}\left\{\left\|\mathbf{p}_u\right\|^2\right\}
    =
    \sum_{j\neq u}\mathbb{E}\left\{\left\|\mathbf{x}_j\right\|^2\right\}
    =
    \sum_{j\neq u}\mathbb{E}\left\{\left\|\mathbf{\Gamma}_j\mathbf{d}_j\right\|^2\right\}
    =
    \sum_{j\neq u}|\mathcal{S}_j|E_j,
\end{equation}
where $E_j$ is the average symbol energy on each allocated chirp subcarrier of user $j$. The average leakage energy per unallocated chirp subcarrier of user $u$ is
\begin{equation}
    \bar {E}_{p,u}
    =
    \frac{\mathbb{E}\{\left\|\mathbf{p}_u\right\|^2\}}{N-|\mathcal{S}_u|}
    =
    \frac{\sum_{j\neq u}|\mathcal{S}_j|E_j}{N-|\mathcal{S}_u|}.
\end{equation}
Since $\mathbf{p}_u$ does not overlap with $\mathbf{x}_u$, it can be interpreted as an equivalent unknown signal confined to the chirp subcarriers not allocated to user $u$, with an average energy of $\bar {E}_{p,u}$ per chirp subcarrier. In the special case of equal symbol energy and full loading, i.e., $E_u=E$ for all $u$ and $\sum_{u=1}^U |\mathcal{S}_u| = N$, $\bar {E}_{p,u}=E$ and the aggregate signal $\mathbf{x}_u+\mathbf{p}_u$ of user $u$ has the same covariance as a full-grid single-user AFDM benchmark where all chirp subcarriers are occupied and share equal symbol energy.

To further characterize the interference under practical channel equalization $\mathbf{W}_u$ in \eqref{equ:equlaized}, we retain the noise term and define $\mathbf{G}_u=\mathbf{W}_u\bar{\mathbf{H}}_u$, $\mathbf B_u=\mathbf\Gamma_u^H\mathbf G_u\mathbf\Gamma_u$, and $\mathbf{C}_{u,j}=\mathbf{\Gamma}_u^\mathrm{H}\mathbf{G}_u\mathbf{T}_{u,j}\mathbf{\Gamma}_j$. The demapped signal is given by
\begin{equation}
\begin{aligned}
    \hat{\mathbf d}_u
    =
    \mathbf\Gamma_u^H\mathbf{W}_u\mathbf{y}_u
    =
    \mathbf B_u\mathbf d_u
    +
    \sum_{j\neq u}\mathbf C_{u,j}\mathbf d_j
    +
    \mathbf\Gamma_u^H\mathbf W_u\tilde{\mathbf w}_u.
\end{aligned}
\end{equation}
The output SINR \cite{sui2026mimo} of the $i$-th data symbol of user $u$ is
\begin{equation}\label{equ:SINR}
\begin{aligned}
\gamma_u(i)
&=
\frac{E_u|\mathbf B_u[i,i]|^2}
{I_u^{\mathrm{ICI}}(i)+I_u^{\mathrm{IUI}}(i)
+\sigma_u^2
\|(\mathbf\Gamma_u^H\mathbf W_u)[i,:]\|^2},\\
I_u^{\mathrm{ICI}}(i)
&=
E_u\!\left(
\|\mathbf B_u[i,:]\|^2-|\mathbf B_u[i,i]|^2
\right),\\
I_u^{\mathrm{IUI}}(i)
&=
\sum\nolimits_{j\neq u}E_j\|\mathbf C_{u,j}[i,:]\|^2,
\end{aligned}
\end{equation}
where $i\in\{0,\ldots,N_u-1\}$. The denominator contains three terms, including the inter-carrier interference (ICI) $I_{u}^{\text{ICI}}(i)$ arising from the off-diagonal entries of $\mathbf B_u$, the IUI $I_{u}^{\text{IUI}}(i)$ resulting from cross-basis leakage, and the noise term.

Under ideal ZF equalization, i.e., $\mathbf{G}_u=\mathbf{I}_N$, the leakage-free condition yields $\mathbf{B}_{u}=\mathbf{I}_{N_u}$ and $\mathbf{C}_{u,j}=\mathbf{0}_{N_u\times N_j}$ for any user $j\neq u$. Hence, both ICI and IUI vanish. For a general equalizer, off-diagonal entries of $\mathbf G_u$ may couple cross-basis leakage onto the allocated chirp subcarriers and introduce output IUI. However, for equal symbol energy with the same channel, noise power, and equalizer, the output SINR on each allocated chirp subcarrier is no lower than that at the corresponding index in the full-grid single-user AFDM benchmark.

\section{Residue-Class Resource Allocation}
\label{section_IV}
Based on the sufficient residue-class orthogonality rule derived in Section \ref{Section_III}, we first formulate the chirp-subcarrier allocation as an optimization problem that maximizes the total number of allocated chirp subcarriers in this section. Then, we exploit the residue-class structure to reduce the problem dimension and develop a two-stage allocation framework.

\subsection{Chirp-Subcarrier-Level Resource Allocation Formulation}
\textit{Theorem~\ref{theorem1}} converts the transmitter-side leakage-free design principle into a sufficient residue-class orthogonality rule, expressed as pairwise modular constraints on the allocated chirp subcarrier sets $\{\mathcal S_u\}_{u=1}^{U}$. Based on these constraints, we formulate a chirp-subcarrier-level problem to maximize the total number of allocated chirp subcarriers:
\begin{subequations}\label{prob:carrier_level_allo}
\begin{align}
\max_{\mathcal{S}_u} \,\,\,\,
& \sum_{u=1}^{U} |\mathcal{S}_u|
\label{prob:carrier_level_allo_obj} \\[1mm]
\text{s.t.}\,\,\,\,
& r\not\equiv q \pmod{\widetilde{D}_{u,j}},
\forall u<j,
\forall r\in\mathcal{S}_u,
\forall q\in\mathcal{S}_j,
\label{prob:carrier_level_allo_conflict} \\
& |\mathcal{S}_u| \geq M_u,
\quad \forall u,
\label{prob:carrier_level_allo_demand}
\end{align}
\end{subequations}
where $M_u\in\mathbb{N}$ denotes the minimum number of chirp subcarriers required by user $u$. The objective in \eqref{prob:carrier_level_allo_obj} is to maximize the total number of allocated chirp subcarriers to enhance resource utilization. Constraint~\eqref{prob:carrier_level_allo_conflict} follows the sufficient residue-class orthogonality rule in \textit{Theorem~\ref{theorem1}}, and cardinality  constraint~\eqref{prob:carrier_level_allo_demand} guarantees the minimum resource requirement for each user.

Constraint \eqref{prob:carrier_level_allo_conflict} 
prevents a chirp subcarrier from being allocated to more than one user. Hence, the chirp subcarrier sets $\{\mathcal S_u\}_{u=1}^{U}$ are pairwise disjoint and the objective in \eqref{prob:carrier_level_allo_obj} is upper-bounded by
\begin{equation}
    \sum_{u=1}^{U} |\mathcal{S}_u|\leq N.
\end{equation}
A feasible allocation achieving this upper bound is called a full-loading allocation, in which each of the $N$ chirp subcarriers is allocated to exactly one user, i.e., 
\begin{equation}
    \bigcup_{u=1}^{U}\mathcal{S}_u=\mathbb Z_N, \quad \mathcal{S}_u\cap\mathcal{S}_j=\varnothing, \forall u\neq j.
\end{equation}

\subsection{Residue-Class-Based Resource Allocation Formulation}

Problem~\eqref{prob:carrier_level_allo} directly allocates $N$ chirp subcarriers among $U$ users under numerous chirp-subcarrier-level conflict constraints. Since these constraints depend only on the residue indices under the pairwise moduli $\widetilde D_{u,j}$, many chirp subcarriers have identical conflict relationships, enabling a residue-class-level reduction.

As $\widetilde{D}_{u,j}=\widetilde{D}_{j,u}$, we define the common fine modulus as
\begin{equation}
    L=\max  \{\widetilde{D}_{u,j}:\forall u < j\}.
\end{equation}
Since $N=2^n$, every effective pairwise modulus $\widetilde D_{u,j}$ is a power-of-two divisor of $N$ and hence
\begin{equation}\label{equ:D_mod_L}
    \widetilde D_{u,j}\mid L, \quad \forall u \neq j.
\end{equation}
The $N$ chirp subcarriers are therefore partitioned into $L$ disjoint fine residue classes, each containing $N/L$ chirp subcarriers, given by
\begin{equation}\label{equ:fine_residue_class}
    \mathcal{R}_g^{(L)}
    =
    \left\{
    g+tL\mid
    t=0,1,\ldots,{N}/{L}-1
    \right\},
\end{equation}
where $g=0,\dots,L-1$. Based on this decomposition, we present the following theorem.

\begin{theorem}
\label{theorem2}
In every optimal solution to Problem~\eqref{prob:carrier_level_allo}, each fine residue class $\mathcal R_g^{(L)}$ is either unallocated or entirely allocated to one user.
\end{theorem}

\textit{Proof:} Consider an optimal solution of Problem \eqref{prob:carrier_level_allo} in which
$r\in\mathcal R_g^{(L)}$ is allocated to user $u$. For any other $q\in\mathcal{R}_g^{(L)}$, we have $r-q\equiv0\pmod{L}$. According to \eqref{equ:D_mod_L}, we then have 
\begin{equation}\label{equ:theorem_2_proof}
    r\equiv q\pmod{\widetilde{D}_{u,j}},\quad\forall j\neq u.
\end{equation}
Therefore, $q$ cannot be allocated to another user $j\neq u$, since this would violate the conflict constraint in \eqref{prob:carrier_level_allo_conflict}.

Next, consider any chirp subcarrier $s$ allocated to another user $j\neq u$ in this optimal solution. Since the solution is feasible and $r$ is allocated to user $u$, we have $r\not\equiv s\pmod{\widetilde D_{u,j}}$. According to \eqref{equ:theorem_2_proof}, we then have
\begin{equation}
    q\not\equiv s\pmod{\widetilde{D}_{u,j}}.
\end{equation}
Hence, if $q$ were unallocated, assigning it to user $u$ would preserve feasibility while increasing the objective in \eqref{prob:carrier_level_allo_obj}, which contradicts optimality. Therefore, $q$ must already be allocated to user $u$. Since $q$ is arbitrary, $\mathcal R_g^{(L)}$ is entirely allocated to user $u$. \hfill$\blacksquare$

According to \textit{Theorem \ref{theorem2}}, Problem~\eqref{prob:carrier_level_allo} can be reduced without loss of optimality. Specifically, for $r\in\mathcal{R}_g^{(L)}$ and $q\in\mathcal{R}_f^{(L)}$, there exist $t,s\in\left\{0,1,\ldots,N/L-1\right\}$ such that $r=g+tL$ and $q=f+sL$. According to \eqref{equ:D_mod_L}, we then have
\begin{equation}
    r-q
    \equiv
    g-f
    \pmod{\widetilde D_{u,j}},\quad\forall u\neq j.
\end{equation}
Therefore, for all $r\in\mathcal R_g^{(L)}$,
$q\in\mathcal R_f^{(L)}$, and $u\neq j$, we have
\begin{equation}\label{equ:carrier_group_conflict_equivalence}
    r\not\equiv q\pmod{\widetilde D_{u,j}}
    \quad\Longleftrightarrow\quad
    g\not\equiv f\pmod{\widetilde D_{u,j}}.
\end{equation}
Hence, the chirp-subcarrier-level conflict constraints \eqref{prob:carrier_level_allo_conflict} between $r\in\mathcal{R}_g^{(L)}$ and $q\in\mathcal{R}_f^{(L)}$ can be represented exactly by the fine residue-class indices $g$ and $f$.

Let $\mathbb{Z}_L=\{0,1,\ldots,L-1\}$ denote the set of fine residue-class indices and let $\mathcal{I}_{u}\subseteq\mathbb{Z}_L$ denote the fine residue-class index set allocated to user $u$. The corresponding chirp-subcarrier set is
\begin{equation}
    \mathcal{S}_u=\bigcup_{g\in\mathcal{I}_{u}}\mathcal{R}_g^{(L)}.
\end{equation}

Based on \textit{Theorem~\ref{theorem2}} and \eqref{equ:carrier_group_conflict_equivalence}, Problem~\eqref{prob:carrier_level_allo} can be reduced to a residue-class-level optimization problem
\begin{subequations}\label{prob:residue_class_allo}
\begin{align}
\max_{\mathcal{I}_{u}} \,\,
& \frac{N}{L}\sum_{u=1}^{U} |\mathcal{I}_{u}|
\label{prob:residue_class_allo_obj} \\[1mm]
\text{s.t.}\,\,
& g\not\equiv f \pmod{\widetilde{D}_{u,j}},
\forall u<j,
\forall g\in\mathcal{I}_u,
\forall f\in\mathcal{I}_j,
\label{prob:residue_class_allo_conflict} \\
& |\mathcal{I}_u| \geq \eta_u,
\quad \forall u,
\label{prob:residue_class_allo_demand}
\end{align}
\end{subequations}
where $\eta_u$ denotes the minimum number of fine residue classes required by user $u$ and is defined as
\begin{equation}\label{equ:eta_u}
    \eta_u
    =
    \left\lceil
    \frac{M_u}{N/L}
    \right\rceil
    =
    \left\lceil
    \frac{M_uL}{N}
    \right\rceil.
\end{equation}

Similarly, constraint \eqref{prob:residue_class_allo_conflict} ensures that each fine residue class can be allocated to at most one user. Hence, the objective of Problem \eqref{prob:residue_class_allo} satisfies
\begin{equation}
    \frac{N}{L}
    \sum_{u=1}^{U}|\mathcal{I}_u|
    \leq N.
\end{equation}
Accordingly, a feasible allocation is full-loading if and only if each fine residue class is allocated to exactly one user, i.e., 
\begin{equation}\label{equ:full_loading_res}
    \bigcup_{u=1}^{U}\mathcal{I}_u=\mathbb Z_L, \quad \mathcal{I}_u\cap\mathcal{I}_j=\varnothing, \forall u\neq j,
\end{equation}
in which case the objective reaches its upper bound $N$.

Problem~\eqref{prob:residue_class_allo} reduces the number
of elementary allocation units from $N$ chirp subcarriers to $L$ fine residue classes. When $L=N$, each fine residue class contains only one chirp subcarrier, and Problem~\eqref{prob:residue_class_allo} coincides with Problem~\eqref{prob:carrier_level_allo}.

Rather than directly solving this optimization problem, we propose a two-stage allocation framework:

\begin{itemize}
    \item{Each user $u$ is allocated exactly $\eta_u$ fine residue classes while satisfying constraint \eqref{prob:residue_class_allo_conflict}, forming a representative allocation. Such an allocation induces a feasible but generally non-optimal solution to Problem \eqref{prob:residue_class_allo}.}
    \item{The remaining fine residue classes are assigned through a deterministic extension rule based on the representative allocation. As will be proved in Subsection~\ref{subsection_IV_D}, any feasible representative allocation can be extended to a feasible full-loading allocation that attains the objective upper bound of Problem~\eqref{prob:residue_class_allo}.}
\end{itemize}

\subsection{Representative Allocation and Recursive Search}
Let $\mathcal A_u\subseteq\{0,1,\ldots,L-1\}$ denote the representative set of user $u$. The first-stage representative allocation problem is
\begin{subequations}\label{prob:representative_allo}
\begin{align}
\text{find}\,\,
& \{\mathcal{A}_u\}_{u=1}^{U},
\\
\text{s.t.}\,\,
& g\not\equiv f \pmod{\widetilde{D}_{u,j}},
\forall u<j,
\forall g\in\mathcal{A}_u,
\forall f\in\mathcal{A}_j,
\label{prob:representative_allo_conflict}
\\
& |\mathcal{A}_{u}|=\eta_u,\quad \forall u.
\label{prob:representative_allo_cardinality}
\end{align}
\end{subequations}
Assigning a representative set to one user may render some fine residue classes unavailable to the remaining users under the pairwise modular constraints \eqref{prob:representative_allo_conflict}. Hence, we develop a user-by-user recursive search algorithm to find a feasible representative allocation.

Let $\mathcal U$ denote the set of users whose representative sets have not yet been assigned, and let $\mathcal{C}_u$ denote the current candidate set of fine residue classes available to user $u$, initialized as
\begin{equation}
    \mathcal U=\{1,2,\ldots,U\},\quad
    \mathcal C_u=\{0,1,\ldots,L-1\},
    \quad \forall u.
\end{equation}
For an unassigned user $u$, any subset $\mathcal{T}_u\subseteq \mathcal{C}_u$ satisfying $|\mathcal{T}_u|=\eta_u$ is referred to as a candidate representative set. The number of possible candidate representative sets for user $u$ is $\binom{|\mathcal C_u|}{\eta_u}$, and each of them may initiate a different search branch.

At each recursion level with $\mathcal U\neq\varnothing$, an unassigned user $u^{\star}\in\mathcal{U}$ is selected according to
\begin{equation}\label{equ:dynamic_user_selection}
    u^{\star}
    =
    \arg\min_{u\in\mathcal U} \binom{|\mathcal C_u|}{\eta_u},
\end{equation}
with ties broken by the smaller user index. This selection rule minimizes the number of possible search branches at the current recursion level. 

After selecting $u^\star$, let $\mathcal U'=\mathcal U\setminus\{u^\star\}$ denote the set of remaining users. We evaluate the element destructiveness of each candidate fine residue class $g\in\mathcal{C}_{u^\star}$ as
\begin{equation}\small\label{equ:single_destructiveness}
\begin{aligned}
    \Delta_{u^\star}
    \left(
    g
    \right)
    =
    \sum_{j\in\mathcal U'}
    \Bigl|
    \Bigl\{
    f\in\mathcal C_j:
    \, f\equiv g
    \pmod{\widetilde D_{j,u^\star}}
    \Bigr\}
    \Bigr|.
\end{aligned}
\end{equation}
It counts the candidate fine residue classes removed from the current candidate sets of all remaining users if $g$ were assigned to user $u^\star$. The elements of $\mathcal{C}_{u^\star}$ are arranged in non-decreasing order of their element destructiveness, forming the ordered candidate list $\bar{\mathcal{C}}_{u^\star}=\bigl[\bar{g}_0,\bar{g}_1,\dots,\bar{g}_{|\mathcal{C}_{u^\star}|-1}\bigr]$ that satisfies
\begin{equation}
    \Delta_{u^\star}(\bar{g}_0)\leq\Delta_{u^\star}(\bar{g}_1)\leq\dots\leq\Delta_{u^\star}(\bar{g}_{|\mathcal{C}_{u^\star}|-1}).
\end{equation}
Ties are broken by the smaller fine-residue-class index. This ordering is computed once at each recursion level and remains fixed during the enumeration of the candidate representative sets of user $u^{\star}$.

Subsequently, each candidate representative set is represented as $\mathcal{T}_{u^\star}=\{\bar g_{i_1},\ldots,\bar g_{i_{\eta_{u^\star}}}\}$, where $0\leq i_1<\cdots<i_{\eta_{u^\star}}
\leq|\mathcal C_{u^\star}|-1$. The corresponding index tuples are enumerated on demand in lexicographic order. Each candidate representative set defines one search branch. For each branch, the candidate set of every remaining user $j\in\mathcal U'$ is updated according to \eqref{prob:representative_allo_conflict} as
\begin{equation}\label{equ:candidate_set_update}
    \mathcal C_j'
    =
    \left\{
    f\in\mathcal C_j:
    f\not\equiv g\pmod{\widetilde D_{j,u^\star}},
    \ \forall g\in\mathcal T_{u^\star}
    \right\}.
\end{equation}

If $|\mathcal C_j'|\geq\eta_j$ for every $j\in\mathcal U'$, the current search branch may still be feasible. The algorithm then recursively searches for the representative allocation of the remaining users with updated fine-residue-class candidate sets $\{\mathcal C_j'\}_{j\in\mathcal U'}$.

If $|\mathcal C_j'|<\eta_j$ for any $j\in\mathcal U'$, the current search branch is not feasible and is discarded without a recursive call. Furthermore, if the cardinality conditions are satisfied but the subsequent recursive call returns failure, the current candidate representative set is also rejected. In either case, the algorithm examines the next untried candidate representative set according to the lexicographic order.

If the recursive call succeeds, the returned representative allocation of the remaining users is combined with $\mathcal A_{u^\star}=\mathcal T_{u^\star}$ and returned to the preceding recursion level. If no candidate representative set of user $u^\star$ leads to success, the current user-level recursion returns failure to the preceding level.

The recursion terminates successfully when no remaining users need to be assigned, i.e., $\mathcal U=\varnothing$. The recursive search returns the empty remaining allocation, and the complete representative allocation is formed by combining the representative sets obtained along the recursive search path. 

\textbf{Algorithm}~\ref{alg:recursive_rep_allocation} summarizes the proposed user-by-user recursive search procedure. It is a complete search for Problem~\eqref{prob:representative_allo}. The user-selection rule and the element-destructiveness ordering are employed as branch-ordering heuristics to improve search efficiency. The recursive search terminates once the first feasible representative allocation is found and returns failure only after all candidate search branches have failed.

\begin{algorithm}
\caption{Recursive Representative Allocation}
\label{alg:recursive_rep_allocation}
\begin{algorithmic}[1]
\STATE \textbf{Input:} Common fine modulus $L$, demands
$\{\eta_u\}_{u=1}^{U}$, and pairwise moduli
$\{\widetilde D_{u,j}\}_{u\neq j}$
\STATE \textbf{Output:} Feasible representative allocation
$\{\mathcal A_u\}_{u=1}^{U}$, or failure

\STATE \textbf{Recursive search function}
$\textsc{Search}
(\mathcal U,\{\mathcal C_u\}_{u\in\mathcal U})$\textbf{:}
\IF{$\mathcal U\neq\varnothing$}
    \STATE Select $u^\star$ according to \eqref{equ:dynamic_user_selection}, with ties broken by the smaller user index
    \STATE $\mathcal U'\leftarrow
    \mathcal U\setminus\{u^\star\}$
    \STATE Arrange $\mathcal C_{u^\star}$ into the ordered candidate list $\bar{\mathcal C}_{u^\star}$ in non-decreasing
    order of \eqref{equ:single_destructiveness}, with ties broken by the smaller fine-residue-class index
    \FOR{each candidate representative set $\mathcal T_{u^\star}$ generated on demand from $\bar{\mathcal C}_{u^\star}$ in lexicographic order}
        \FOR{each $j\in\mathcal U'$}
            \STATE Update $\mathcal C_j'$ according to
            \eqref{equ:candidate_set_update}
        \ENDFOR
        \IF{$|\mathcal C_j'|\geq\eta_j,\ \forall j\in\mathcal U'$}
            \STATE $\mathcal A^{\rm res}\leftarrow
            \textsc{Search}
            (\mathcal U',
            \{\mathcal C_j'\}_{j\in\mathcal U'})$
            \IF{$\mathcal A^{\rm res}\neq\mathrm{failure}$}
                \STATE \textbf{return}
                $\mathcal A^{\rm res}
                \cup
                \{\mathcal A_{u^\star}
                =
                \mathcal T_{u^\star}\}$
            \ENDIF
        \ENDIF
    \ENDFOR
\STATE \textbf{return} failure
\ELSE
    \STATE \textbf{return} $\varnothing$ (Successful terminal state)
\ENDIF

\STATE \textbf{Initial call:}
\STATE Let $\mathcal U=\{1,\ldots,U\}$, $\mathcal C_u=\{0,\ldots,L-1\},\forall u$
\STATE \textbf{return}
$\textsc{Search}
(\mathcal U,\{\mathcal C_u\}_{u\in\mathcal U})$

\end{algorithmic}
\end{algorithm}

\begin{remark}
    The feasibility of Problem \eqref{prob:representative_allo} provides the following parameter-design guidelines in a downlink heterogeneous-$c_1$ AFDMA system:
    \begin{itemize}
        \item{\emph{Cardinality condition:} Since each fine residue class can be assigned to at most one user, a necessary condition is $\sum_{u=1}^{U}\eta_u \leq L$. Otherwise, constraint \eqref{prob:representative_allo_cardinality} cannot be satisfied.}
        
        \item{\emph{Pairwise-modulus condition:} Every user pair must satisfy $\widetilde D_{u,j}\geq 2$. If $\widetilde D_{u,j}=1$, all fine residue classes are congruent modulo $\widetilde D_{u,j}$, making \eqref{prob:representative_allo_conflict} infeasible for users $u$ and $j$. Such post-chirp parameter pairs should therefore be excluded. A typical example is the coexistence of OFDM $(c_1=0)$ and OCDM $(c_1=1/(2N))$, yielding $|K_{u,j}|=1$ and $\widetilde D_{u,j}=1$, so no leakage-free residue-class allocation exists for this pair.}
    \end{itemize}

\end{remark}

\subsection{Full-Loading Extension and Optimality}
\label{subsection_IV_D}

Given a feasible representative allocation $\{\mathcal{A}_u\}_{u=1}^{U}$ from the first stage, the second stage assigns the remaining fine residue classes through a deterministic extension rule. Let
\begin{equation}
    \mathcal{Q}
    =
    \mathbb{Z}_L 
    \setminus 
    \bigcup_{u=1}^U \mathcal{A}_u
\end{equation}
denote the set of remaining unassigned fine residue classes after representative allocation, and define the set of relevant moduli as
\begin{equation}\label{equ:relevant_moduli}
    \mathcal{D} = \{1\}\cup\{\widetilde D_{u,j}:\forall u<j\}.
\end{equation}
For any $a,b\in\mathbb{Z}_L$, define
\begin{equation}\label{equ:rho_ab}
    \rho(a,b) = \max \{D\in\mathcal{D}:a\equiv b\pmod D\}.
\end{equation}
The quantity $\rho(a,b)$ characterizes the largest modulus $D\in\mathcal{D}$ under which $a$ and $b$ belong to the same residue class. A larger $\rho(a,b)$ indicates a higher modular similarity between $a$ and $b$.

For each $b\in\mathcal{Q}$, define its modular similarity to user $u$ as
\begin{equation}\label{equ:rho_ub}
    \rho_u(b)=\max_{a\in\mathcal{A}_u} \rho(a,b),\quad b\in\mathcal{Q}.
\end{equation}
It characterizes the largest modular similarity between $b$ and the representative set of user $u$.

Then each remaining fine residue class $b\in\mathcal{Q}$ is assigned to a user attaining the largest modular similarity, namely,
\begin{equation}\label{equ:phi_b}
    \phi(b)
    =
    \operatorname{TB}\big(\arg\max_{1\leq i\leq U}\rho_i(b)\big),
\end{equation}
where the operator $\operatorname{TB}(\cdot)$ denotes a deterministic tie-breaking operator that selects one user from any nonempty subset of $\{1,\ldots,U\}$. It can be replaced by any deterministic user-priority rule.

Accordingly, the final set of fine residue classes assigned to user $u$ is the union of the representative set and the above extended allocation, i.e.,
\begin{equation}\label{equ:I_u}
    \mathcal I_u
    =
    \mathcal A_u\cup\{b\in\mathcal{Q}:\phi(b)=u\},
    \quad \forall u.
\end{equation}
The validity of this extension rule is established in
\textit{Theorem~\ref{theorem3}}.

\begin{theorem}\label{theorem3}
    For any feasible representative allocation $\{\mathcal A_u\}_{u=1}^{U}$, the extension rule in \eqref{equ:phi_b} and \eqref{equ:I_u} produces a feasible full-loading allocation $\{\mathcal I_u\}_{u=1}^{U}$ for the residue-class-level optimization problem \eqref{prob:residue_class_allo}.
\end{theorem}

\textit{Proof:} See Appendix~\ref{app:theorem3}. \hfill$\blacksquare$

Together with the representative-allocation formulation, \textit{Theorem~\ref{theorem3}} establishes the infeasibility-or-optimal-full-loading property of Problem~\eqref{prob:residue_class_allo}. If no feasible representative allocation exists, Problem~\eqref{prob:residue_class_allo} is also infeasible. Otherwise, \textit{Theorem~\ref{theorem3}} shows that any feasible representative allocation can be extended to a feasible full-loading allocation. Since full loading attains the objective upper bound in Problem~\eqref{prob:residue_class_allo}, this allocation is optimal.

Therefore, solving Problem~\eqref{prob:residue_class_allo} reduces to the recursive search for a representative allocation in Algorithm \ref{alg:recursive_rep_allocation}, followed by the deterministic full-loading extension in \eqref{equ:phi_b} and \eqref{equ:I_u}, which completes the leakage-free chirp subcarrier allocation design for the proposed downlink heterogeneous-$c_1$ system.

\section{Simulation Results}

This section presents numerical search experiments and Monte Carlo simulations to evaluate the proposed downlink heterogeneous-$c_1$ AFDMA design. Unless otherwise stated, the simulation parameters are listed in Table~\ref{tab:simulation_parameters}.
We consider illustrative 3- and 4-user configurations with heterogeneous mobility conditions. Their post-chirp parameters are treated as prescribed user-specific settings, reflecting individual channel and other implementation requirements.
Since the pre-chirp $c_2$ is irrelevant to the leakage-free condition, all AFDM users employ the same $c_2=1/(2N)$. The 4-user system includes an OCDM user as a special AFDM configuration. The user-specific parameters are listed in Table~\ref{tab:user_parameters}.

\begin{table}[!t]
\centering
\caption{Simulation Parameters}
\label{tab:simulation_parameters}
\renewcommand{\arraystretch}{1.05}
\begin{tabular}{c c}
\toprule
\textbf{Parameter} & \textbf{Value} \\
\midrule
Number of chirp subcarriers $N$ & 256 \\
Carrier frequency & 24 GHz \\
Subcarrier spacing $\Delta f$ & 7.5 kHz \\
Channel model & Extended Vehicular A (EVA) \cite{3gpp36104}\\
Modulation scheme & QPSK \\
Equalizer & LC-BMMSE \cite{mao2026joint} \\
Common $c_{2,u}$ & $1/(2N)$ \\
\bottomrule
\end{tabular}
\end{table}

\begin{table}[!t]
\centering
\caption{User Configurations for the 3- and 4-User Cases}
\label{tab:user_parameters}
\renewcommand{\arraystretch}{1.0}
\begin{tabular}{c c c c c c}
\toprule
\textbf{Waveform}
& \makecell[c]{\textbf{3-user}\\\textbf{index}}
& \makecell[c]{\textbf{4-user}\\\textbf{index}}
& \makecell[c]{$v_{u,\max}$\\ \textbf{(km/h)}}
& \makecell[c]{\textbf{Normalized}\\\textbf{Doppler}}
& $c_{1,u}$ \\
\midrule
OCDM & -- & 1 & 5 & 0.015
& $1/(2N)$ \\
AFDM & 1 & 2 & 50 & 0.15
& $3/(2N)$ \\
AFDM & 2 & 3 & 450 & 1.33
& $9/(2N)$ \\
AFDM & 3 & 4 & 900 & 2.67
& $17/(2N)$ \\
\bottomrule
\end{tabular}
\end{table}

\subsection{Residue-Class Allocation Results}
The computational burden of residue-class allocation arises mainly from the representative-allocation stage. Algorithm~\ref{alg:recursive_rep_allocation} combines the proposed user-selection (SU) rule with element-destructiveness (DE) ordering to improve its practical search efficiency. To evaluate this proposed method, we compare it with brute-force enumeration (BF) and three recursive-search configurations: fixed user with fixed element ordering (FU-FE), fixed user with the proposed element destructiveness ordering (FU-DE) and the proposed user selection with fixed element ordering (SU-FE). Under FU, users are selected in ascending index order. Under FE, the elements of the candidate set $\mathcal C_{u^\star}$ retain their ascending index order, and the candidate representative sets $\mathcal T_{u^\star}$ are enumerated lexicographically. BF enumerates the candidate representative allocations lexicographically without candidate-set updating or pruning.

Define the minimum-demand load rate as $\lambda={\sum_{u=1}^{U}\eta_u}/{L}$. For each $\lambda$, we enumerate all demand vectors $\{\eta_1,\dots,\eta_U\}$ satisfying $\eta_u\geq1$, $\forall u$ and $\sum_{u=1}^{U}\eta_u=\lambda L$. Feasible and infeasible instances are evaluated separately. For all methods, one search trial is counted whenever a candidate representative set is generated and examined at a search level, and the same counting rule is applied to BF.

\begin{figure}
    \centering

    \begin{subfigure}[b]{0.48\linewidth}
        \centering
        \includegraphics[width=\linewidth]{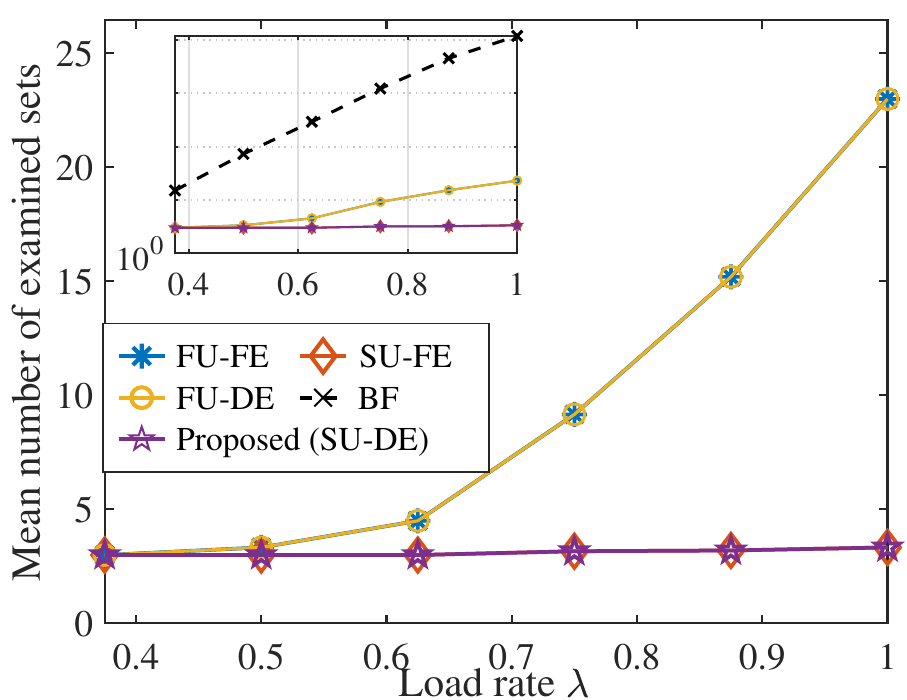}
        \caption{3-user feasible instances}
        \label{fig:3_user_feasible}
    \end{subfigure}
    \hfill
    \begin{subfigure}[b]{0.48\linewidth}
        \centering
        \includegraphics[width=\linewidth]{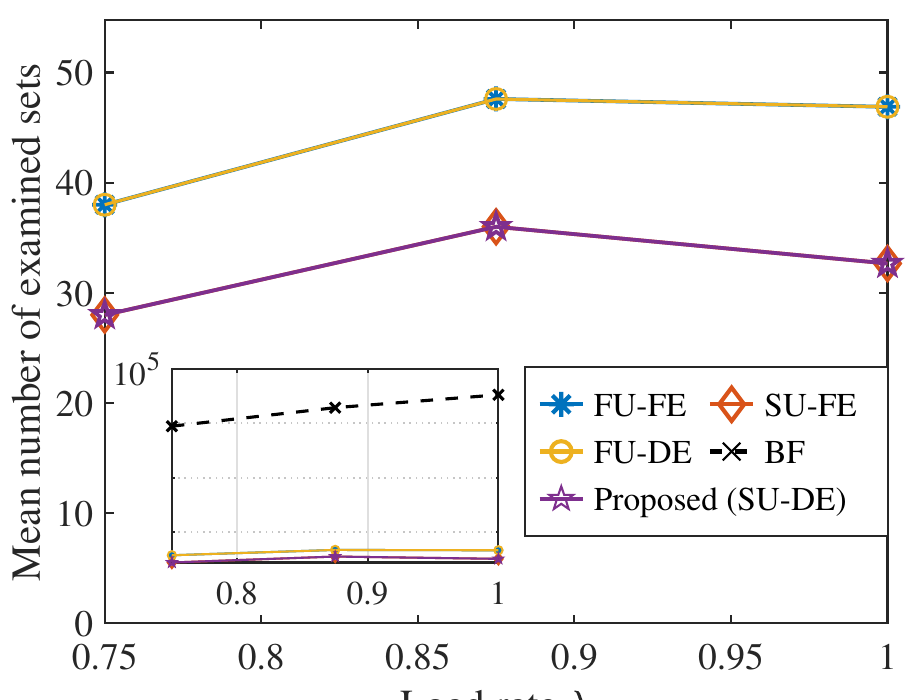}
        \caption{3-user infeasible instances}
        \label{fig:3_user_infeasible}
    \end{subfigure}

    \begin{subfigure}[b]{0.48\linewidth}
        \centering
        \includegraphics[width=\linewidth]{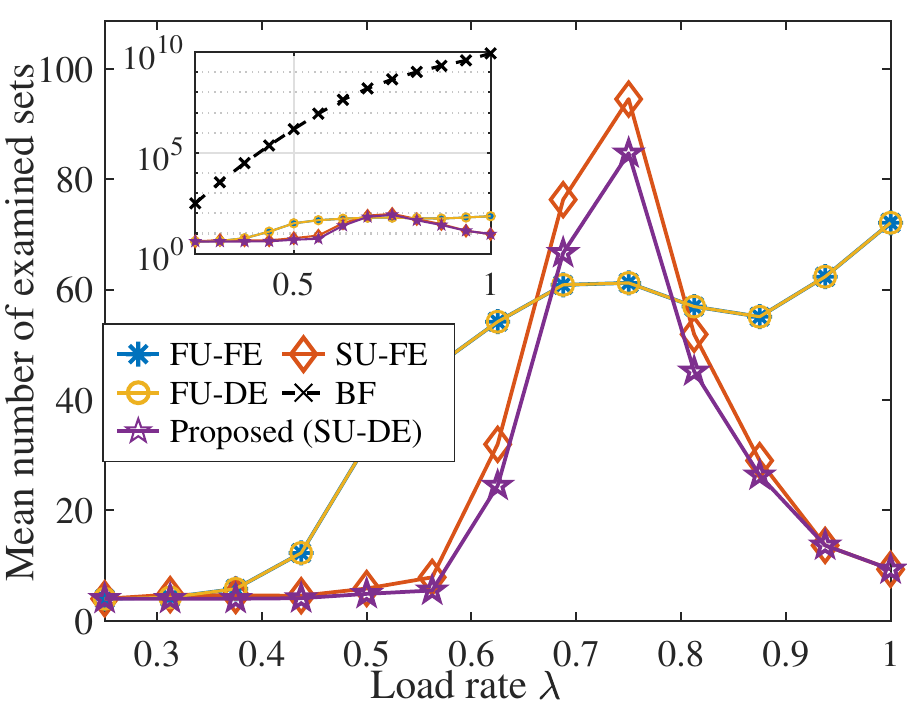}
        \caption{4-user feasible instances}
        \label{fig:4_user_feasible}
    \end{subfigure}
    \hfill
    \begin{subfigure}[b]{0.48\linewidth}
        \centering
        \includegraphics[width=\linewidth]{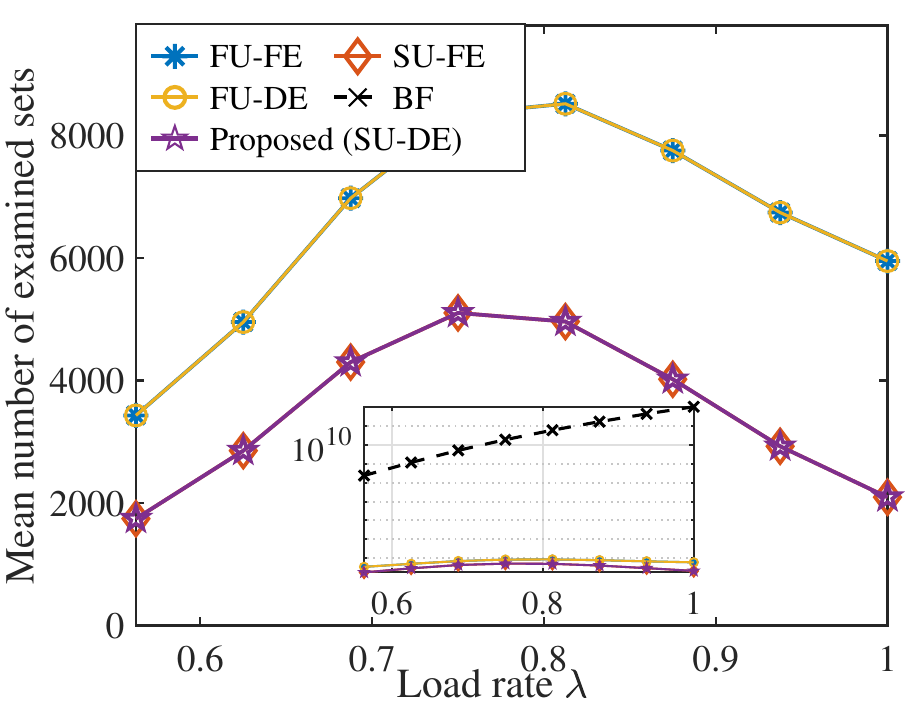}
        \caption{4-user infeasible instances}
        \label{fig:4_user_infeasible}
    \end{subfigure}

    \caption{Search efficiency comparison for the 3-user and 4-user parameter settings listed in Table~\ref{tab:user_parameters}. Each point is averaged over all instances with the same load rate and feasibility status.}
    \label{fig:residue_class_allocation}
\end{figure}

Fig.~\ref{fig:residue_class_allocation} compares the mean number of examined candidate representative sets at each load rate, separately for feasible and infeasible instances. Recursive search outperforms BF by updating candidate sets and pruning infeasible branches. 

For feasible instances in Figs.~\ref{fig:3_user_feasible} and~\ref{fig:4_user_feasible}, the search terminates upon finding the first feasible representative allocation, making the branch order important. Comparing FU-FE with SU-FE shows that the proposed SU rule generally reduces the search count but may increase the count at some load rates in Fig.~\ref{fig:4_user_feasible}. Comparing FU-FE with FU-DE shows that the proposed DE achieves limited improvement under FU. In the proposed method, DE mitigates the degradation caused by unfavorable user selections when combined with SU. Its search-efficiency gain is more pronounced under SU than FU, indicating that the proposed DE rule is more effective after the proposed SU rule.

For infeasible instances in Fig.~\ref{fig:3_user_infeasible} and Fig.~\ref{fig:4_user_infeasible}, FU-FE and FU-DE yield the same mean search counts, as do SU-FE and the proposed method. For these instances, the search process is dominated by infeasibility certification and every branch that is not eliminated by the cardinality constraint must be exhausted. Consequently, DE only reorders branches, giving identical counts compared with FE under the same user-selection strategy. By contrast, SU prioritizes the user with the fewest possible candidate representative sets and tends to expose cardinality constraint violations earlier. Thus, the improvement of the proposed method mainly results from the proposed SU rule for infeasible instances.

Across all tested instances, the proposed method is no worse than FU-FE in more than $93$\% of the cases and strictly better in more than $63$\%. Nevertheless, a few difficult instances may require more examined candidate representative sets and dominate the mean at some load rates.

\subsection{Communication Performance Comparison}

\begin{figure}[htbp!]
    \centering
    \includegraphics[width=1\linewidth]{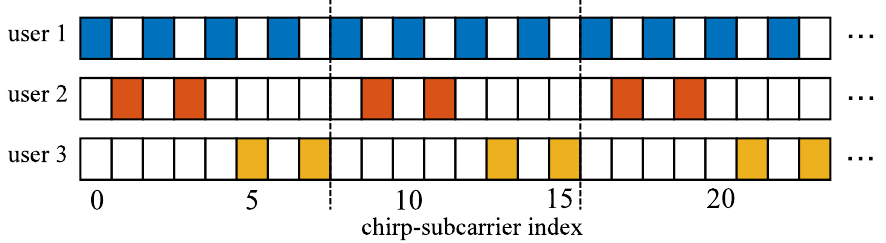}
    \caption{Chirp-subcarrier mapping for the $3$-user allocation in Table~\ref{tab:residue_class_allocation}. Colored and white blocks indicate allocated and unallocated positions for each user, respectively.}
    \label{fig:residue_to_subcarrier}
\end{figure}

For communication performance evaluation, we set $\eta_u = 2$ for all users. Applying the proposed two-stage residue-class allocation to the 3-user and 4-user configurations in Table~\ref{tab:user_parameters} yields the representative and full-loading allocations in Table~\ref{tab:residue_class_allocation}. The full-loading allocation sets $\mathcal{I}_u$ are then mapped to chirp subcarriers, as illustrated in Fig.~\ref{fig:residue_to_subcarrier} using the 3-user system as an example. Perfect CSI is assumed, and each user employs the LC-BMMSE equalizer in \cite{mao2026joint}. Unless otherwise stated, all BER benchmarks use the same integer-delay and fractional-Doppler channel realizations and per-symbol power normalization. The signal-to-noise ratio (SNR) at user $u$ is defined as $\mathrm{SNR}=E_d/\sigma_u^2$, where $E_d=1$ denotes the average symbol energy common to all users and $\sigma_u^2$ is the variance of each complex AWGN sample at user $u$.

\begin{table}[htbp!]
\centering
\caption{Representative and Full-Loading Residue-Class Allocations}
\label{tab:residue_class_allocation}
\renewcommand{\arraystretch}{1.05}
\begin{tabular}{c c c c}
\toprule
\textbf{Case}
& \textbf{User Index}
& \makecell[c]{\textbf{Representative}\\
\textbf{set $\mathcal A_u$}}
& \makecell[c]{\textbf{Full-loading}\\
\textbf{allocation $\mathcal I_u$}}\\
\midrule
\multirow{3}{*}{\makecell[c]{3-user\\$L=8$}}
& 1 & $\{0,2\}$ & $\{0,2,4,6\}$ \\
& 2 & $\{1,3\}$ & $\{1,3\}$ \\
& 3 & $\{5,7\}$ & $\{5,7\}$ \\
\midrule
\multirow{4}{*}{\makecell[c]{4-user\\$L=16$}}
& 1 & $\{0,2\}$  & $\{0,2\}$ \\
& 2 & $\{1,3\}$  & $\{1,3,5,7,9,11,13,15\}$ \\
& 3 & $\{4,6\}$  & $\{4,6,12,14\}$ \\
& 4 & $\{8,10\}$ & $\{8,10\}$ \\
\bottomrule
\end{tabular}
\end{table}

\begin{figure}[htbp!]
    \centering

    \begin{subfigure}[b]{0.48\linewidth}
        \centering
        \includegraphics[width=1\linewidth]{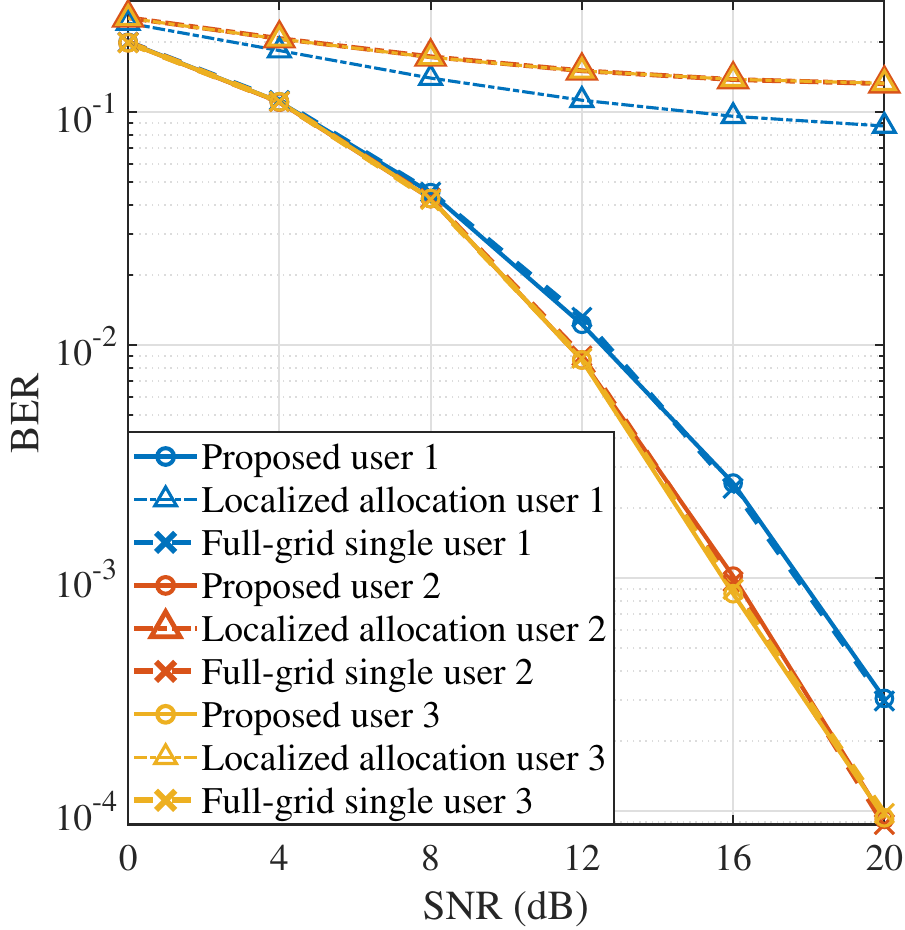}
        \caption{3-user system}
        \label{fig:vs_3user}
    \end{subfigure}
    \hfill
    \begin{subfigure}[b]{0.48\linewidth}
        \centering
        \includegraphics[width=1\linewidth]{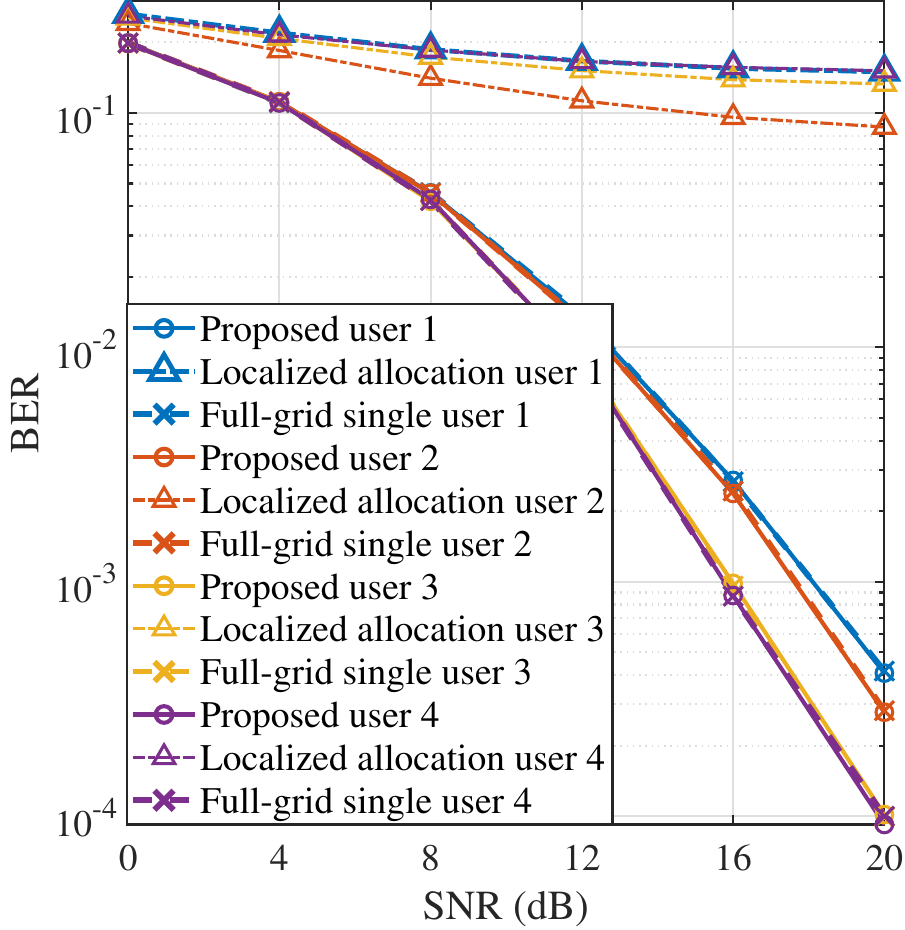}
        \caption{4-user system}
        \label{fig:vs_4user}
    \end{subfigure}

    \caption{Per-user BER comparison of the proposed residue-class and conventional localized allocations in heterogeneous-$c_1$ AFDMA systems, together with the corresponding full-grid single-user AFDM benchmarks.}
    \label{fig:vs_mixed_multi_user}
\end{figure}

Fig.~\ref{fig:vs_mixed_multi_user} compares the per-user BER performance of the heterogeneous-$c_1$ AFDMA systems with the proposed residue-class allocation and the conventional localized DAFT-domain allocation benchmark. The corresponding full-grid single-user AFDM benchmark for each user is also included. For a fair comparison, the two AFDMA schemes use identical chirp parameters and per-user chirp-subcarrier allocation sizes. Following \cite{yin2022pilot,zheng2025partially}, the localized-allocation benchmark assigns user $u$ the contiguous block $\mathcal S_u^{\mathrm{loc}}$ containing $M_u^{\mathrm{loc}}$ chirp subcarriers, expressed as 
$\mathcal S_u^{\mathrm{loc}}
=\{\sum_{k=1}^{u-1}M_k^{\mathrm{loc}},\ldots,
\sum_{k=1}^{u}M_k^{\mathrm{loc}}-1\}$, where $M_u^{\mathrm{loc}}=|\mathcal I_u|N/L$. For the full-grid single-user AFDM benchmark corresponding to user $u$, all other users are silent, while all $N$ chirp subcarriers carry data symbols with the same average symbol energy and chirp parameters $(c_{1,u},c_{2,u})$ as those of user $u$ in the corresponding AFDMA system.

As shown in Fig.~\ref{fig:vs_mixed_multi_user}, the proposed systems avoid the BER floors of the localized-allocation benchmarks. For the considered heterogeneous-$c_1$ configurations, the localized allocation violates the pairwise modular orthogonality conditions, allowing cross-basis leakage to overlap the user-allocated chirp subcarriers. The resulting IUI becomes dominant at high SNR and causes a BER floor. In contrast, the proposed allocation preserves inter-user orthogonality and avoids the BER floors. Moreover, the proposed systems achieve BER close to their corresponding full-grid single-user AFDM benchmarks, consistent with the analysis in Subsection~\ref{subsection_III_D} under equal symbol energy and full loading. The proposed allocation thus supports heterogeneous-$c_1$ transmission without an appreciable BER loss relative to the single-user benchmarks.

\begin{figure}[!htbp]
    \centering
    \includegraphics[width=0.8\linewidth]{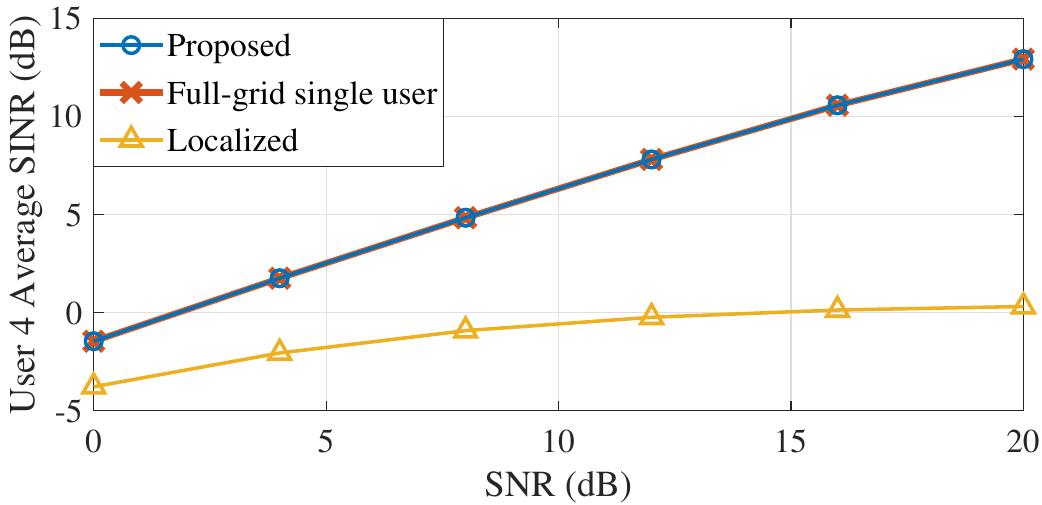}
    \caption{Average output SINR of user $4$ versus SNR in the $4$-user system for the compared methods in Fig.~\ref{fig:vs_mixed_multi_user}.}
    \label{fig:SINR_SNR}
\end{figure}

Fig.~\ref{fig:SINR_SNR} further compares the average output SINR computed from \eqref{equ:SINR} for user $4$ in the $4$-user system, using the same configurations as in Fig.~\ref{fig:vs_mixed_multi_user}. The proposed system achieves SINR close to that of the corresponding full-grid single-user AFDM benchmark and avoids the pronounced IUI-induced SINR ceiling observed with localized allocation.

\begin{figure}[!t]
    \centering

    \begin{subfigure}[b]{0.48\linewidth}
        \centering
        \includegraphics[width=1\linewidth]{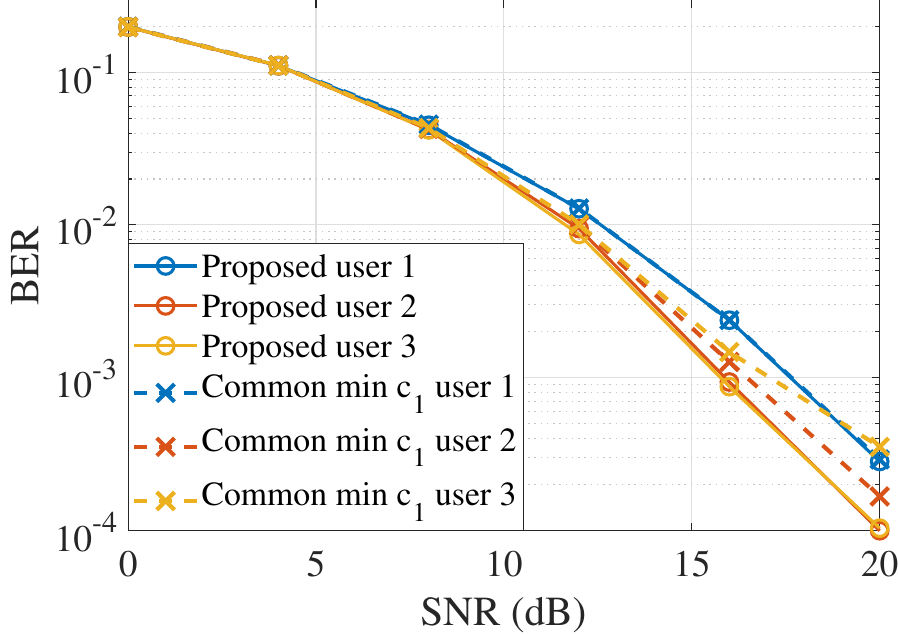}
        \caption{3-user system with common minimum $c_1=3/(2N)$}
        \label{fig:vs_common_min_3user}
    \end{subfigure}
    \hfill
    \begin{subfigure}[b]{0.48\linewidth}
        \centering
        \includegraphics[width=1\linewidth]{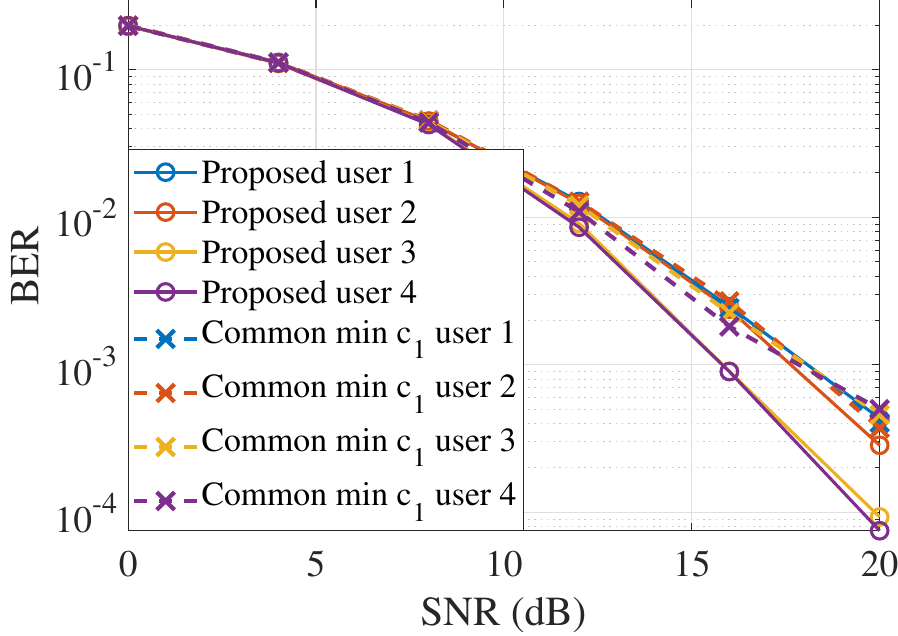}
        \caption{4-user system with common minimum $c_1=1/(2N)$}
        \label{fig:vs_common_min_4user}
    \end{subfigure}

    \begin{subfigure}[b]{0.48\linewidth}
        \centering
        \includegraphics[width=1\linewidth]{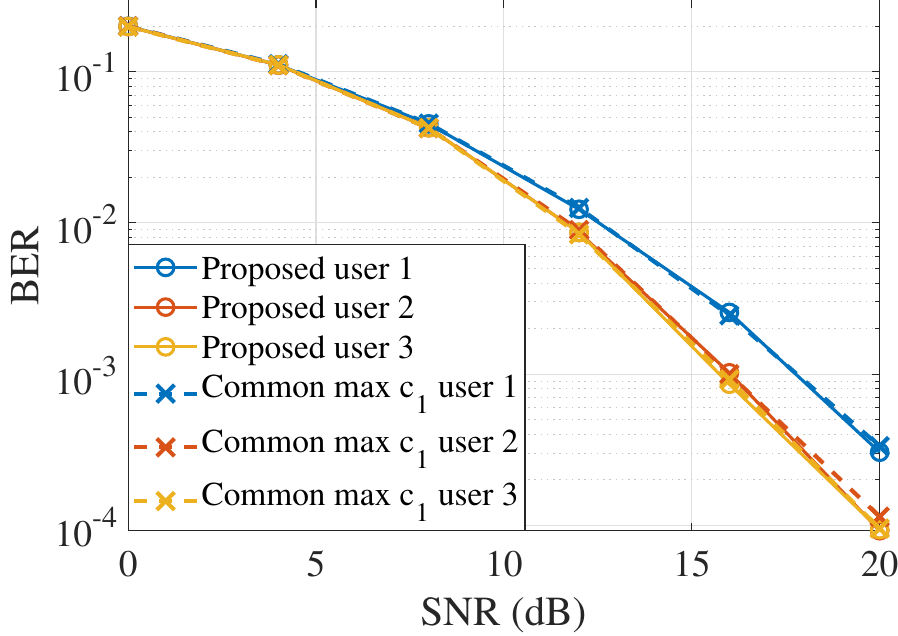}
        \caption{3-user system with common maximum $c_1=17/(2N)$}
        \label{fig:vs_common_max_3user}
    \end{subfigure}
    \hfill
    \begin{subfigure}[b]{0.48\linewidth}
        \centering
        \includegraphics[width=1\linewidth]{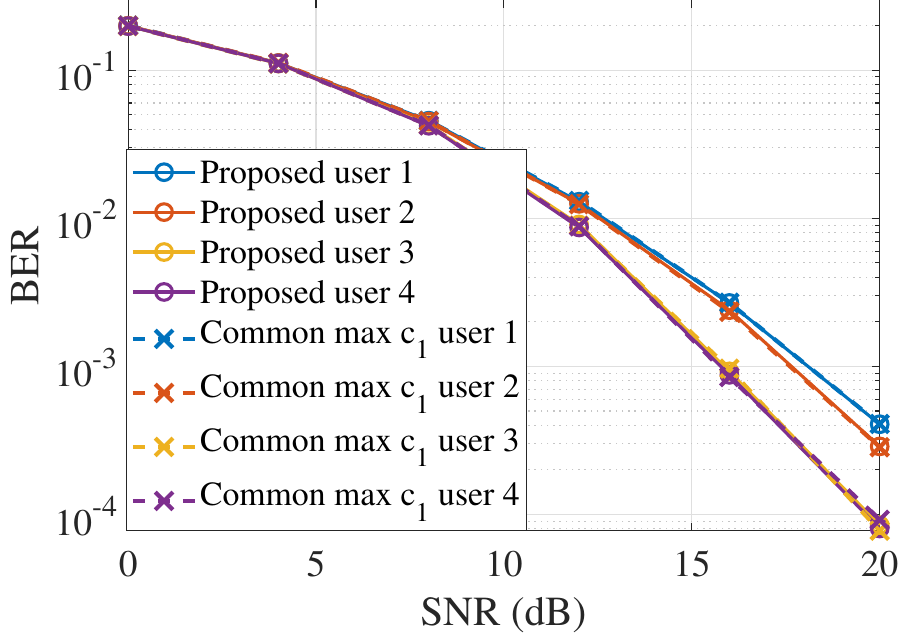}
        \caption{4-user system with common maximum $c_1=17/(2N)$}
        \label{fig:vs_common_max_4user}
    \end{subfigure}

    \caption{Per-user BER comparison between the proposed heterogeneous-$c_1$ AFDMA systems and the conventional common-$c_1$ AFDMA benchmarks.}
    \label{fig:vs_common_multi_user}
\end{figure}

\begin{figure}[!t]
    \centering

    \begin{subfigure}[b]{0.48\linewidth}
        \centering
        \includegraphics[width=1\linewidth]{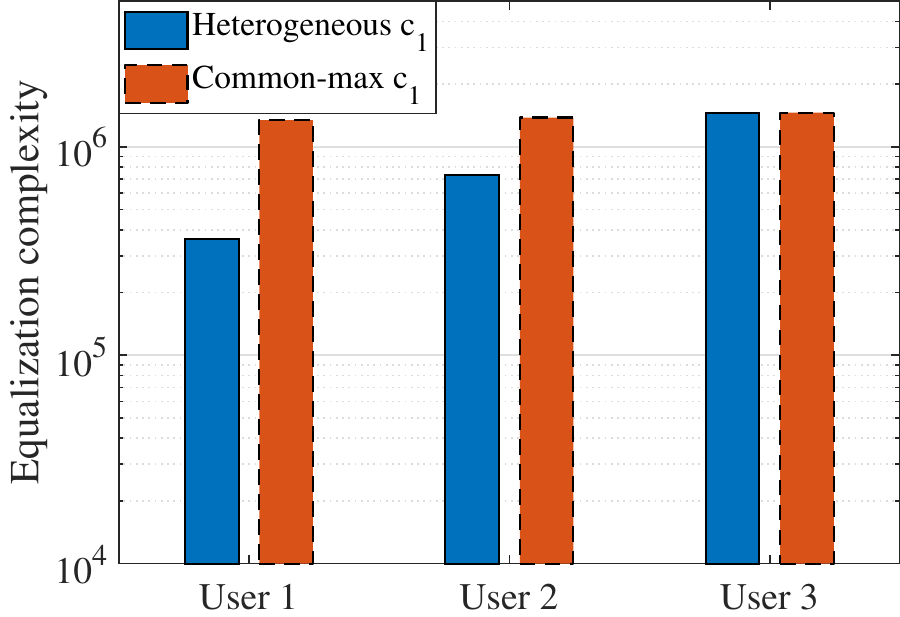}
        \caption{3-user system}
        \label{fig:complexity_3user}
    \end{subfigure}
    \hfill
    \begin{subfigure}[b]{0.48\linewidth}
        \centering
        \includegraphics[width=1\linewidth]{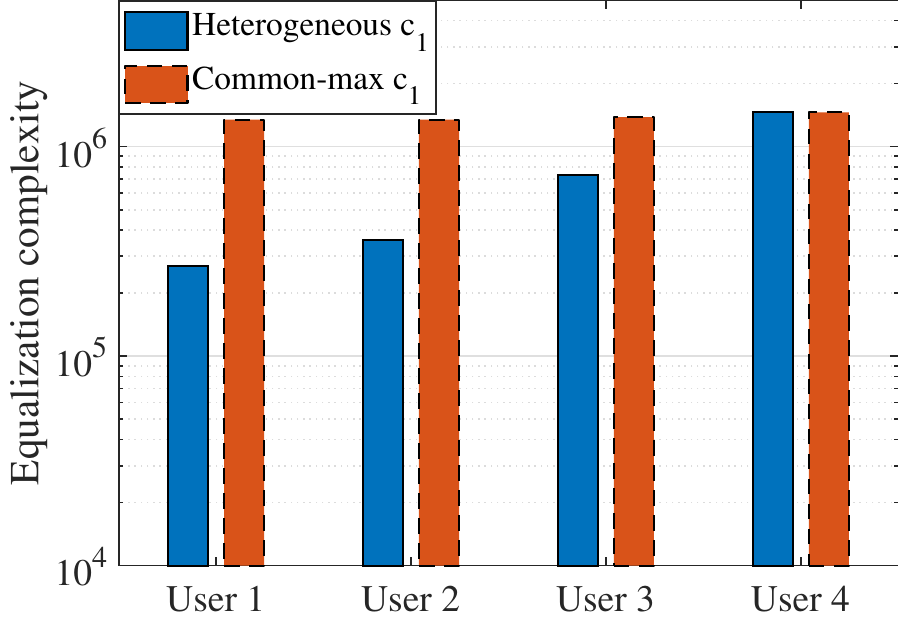}
        \caption{4-user system}
        \label{fig:complexity_4user}
    \end{subfigure}

    \caption{Per-user equalization complexity comparison between the heterogeneous-$c_1$ and common-maximum-$c_1$ AFDMA.}
    \label{fig:complexity_multi_user}
\end{figure}

Fig.~\ref{fig:vs_common_multi_user} compares the per-user BER performance of the proposed heterogeneous-$c_1$ AFDMA systems with common-$c_1$ benchmarks. In each benchmark, all users adopt either the minimum or maximum prescribed $c_1$, retain identical $c_{2}$ values, and use the same disjoint localized allocation blocks as in Fig.~\ref{fig:vs_mixed_multi_user}. The common-minimum-$c_1$ configuration degrades users originally configured with larger $c_1$ values to accommodate higher mobility conditions, particularly users $3$ and $4$ in the 4-user system. 
This is because the shared minimum $c_1$ cannot provide full channel diversity for these users, which require a larger $c_1$ to separate their equivalent channel paths.
By contrast, Figs.~\ref{fig:vs_common_max_3user} and \ref{fig:vs_common_max_4user} show that the common-maximum-$c_1$ configuration does not provide BER improvement, but the receiver complexity is much higher, as shown in Fig.~\ref{fig:complexity_multi_user}. 

Fig.~\ref{fig:complexity_multi_user} compares the per-user receiver complexity of the proposed and common-maximum-$c_1$ systems. For the LC-BMMSE receiver \cite{mao2026joint}, the computational complexity increases with $c_{1,u}$ and can be expressed as $\mathcal{O}(N\left(Nc_{1,u}(l_{u,\mathrm{max}}-l_{u,\mathrm{min}})+k_{u,\mathrm{max}}+w\right)^2)$, where $l_{u,\mathrm{max}}$ and $l_{u,\mathrm{min}}$ are the maximum and minimum integer delay taps, respectively. $k_{u,\mathrm{max}}$ is the maximum integer Doppler tap, and the pruning width parameter $w$ is fixed at $w=30$. As shown in Fig.~\ref{fig:complexity_multi_user}, the common-maximum $c_1$ configuration markedly increases the receiver complexity for users originally assigned smaller $c_{1,u}$ while providing no consistent BER gain. Therefore, the proposed framework achieves BER comparable to the common-maximum-$c_1$ benchmark with lower receiver complexity for users with smaller $c_1$ values. In addition, preserving user-specific waveform configurations offers potential benefits for adapting communication and sensing characteristics to heterogeneous user requirements.

\begin{figure}[!t]
    \centering
    \includegraphics[width=0.8\linewidth]{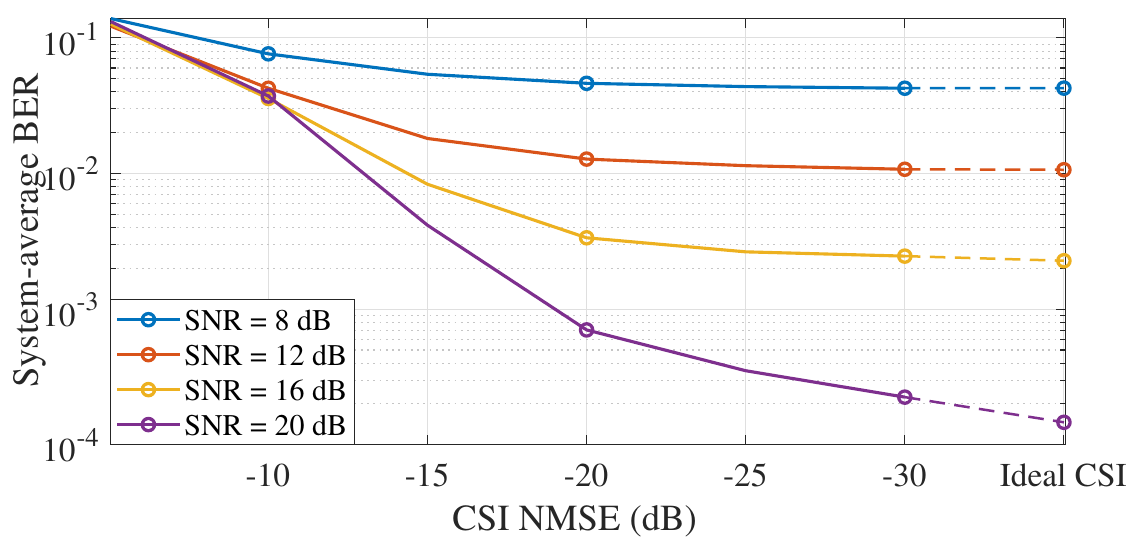}
    \caption{System-average BER versus CSI NMSE for the proposed $4$-user AFDMA system at different SNRs.}
    \label{fig:CSI_NMSE}
    \vspace{-2mm}
\end{figure}

Fig.~\ref{fig:CSI_NMSE} evaluates the effect of imperfect CSI on the proposed method using the $4$-user system as an example. The system-average BER is calculated over all transmitted bits of all users. To model imperfect CSI, the channel matrix employed for equalization is constructed using the channel-gain vector $\hat{\mathbf{h}}_u=\mathbf{h}_u+\mathbf{e}_u$, where $\mathbf{h}_u=[h_{1,u},\ldots,h_{P_u,u}]^{\mathrm T}$ is the actual channel-gain vector and the entries of $\mathbf e_u$ are independent complex Gaussian variables. The CSI error is characterized by $\mathrm{NMSE}=\mathbb E\{\|\mathbf e_u\|^2\}/\mathbb E\{\|\mathbf h_u\|^2\}$. The ``Ideal CSI'' points correspond to $\mathbf{e}_u=\mathbf{0}_{P_u}$. At each SNR, the BER approaches its ideal-CSI value as the NMSE decreases, demonstrating robustness of the proposed system to CSI errors.
\vspace{-1mm}

\section{Conclusion}

This paper developed a downlink AFDMA framework supporting users with heterogeneous post-chirp parameters within the same AFDM block. By analyzing the cross-projection between user-specific DAFT bases, we characterized the cross-basis leakage pattern through quadratic Gauss sums and derived a sufficient pairwise residue-class orthogonality rule to preserve inter-user orthogonality. The chirp-subcarrier allocation problem was reduced to the residue-class level and solved by a two-stage framework comprising a recursive representative-allocation search and a deterministic full-loading extension. We proved that any feasible representative allocation can be extended to an optimal full-loading solution, establishing the infeasibility-or-optimal-full-loading property. The results in this paper establish the theoretical conditions and resource-allocation framework for supporting user-specific waveform configurations in downlink AFDMA, thereby preserving the flexibility to adapt waveform parameters to individual user requirements. Future work will investigate channel estimation and sensing for heterogeneous post-chirp AFDMA, including the joint design of user-specific post-chirp parameters for communication and sensing requirements.
\vspace{-1mm}

\appendices
\section{Proof of Lemma 1}
\label{app:lemma1}
For notational brevity, let $K=K_{u,j}$, $A=r-q$ and $D=D_{u,j}$ in this appendix. Then the quadratic Gauss sum $\mathcal{G}_{u,j}[r,q]$ is written as $\mathcal{G}[A]$, given by
\begin{equation}\label{equ:appendix_g_a}
    \mathcal{G}[A]=\mathcal{G}_{u,j}[r,q]=\sum_{n=0}^{N-1} e^{-\imath2\pi(\frac{A}{N}n+\frac{K}{2N}n^2)}.
\end{equation}
Define
\begin{equation}\label{equ:appendix_g_n}
    g[n]=
    e^{-\imath\frac{2\pi}{N}
    \left(
    A n+\frac{K}{2}n^2
    \right)}.
\end{equation}
Since $N$ is even and $A,K\in\mathbb{Z}$, $g[n]$ is $N$-periodic, i.e., $g[n+N]=g[n]$. 
Using the $N$-periodicity of $g[n]$, we reindex $\mathcal{G}[A]\mathcal{G}^*[A]$ by $h=(n-m)(\mathrm{mod}\,N)$ and substitute \eqref{equ:appendix_g_n} to obtain a further derivation for the squared magnitude of the quadratic Gauss sum in \eqref{equ:appendix_g_a}, shown as
\begin{equation}\label{equ:appendix_G_square_decomposed}
    \begin{aligned}
        |\mathcal{G}[A]|^2
        &=
        \sum_{h=0}^{N-1}\sum_{m=0}^{N-1}
        g[m+h]g^{*}[m]\\
        &=
        \sum_{h=0}^{N-1}
        e^{-\imath\frac{2\pi}{N}
        \left(
        Ah+\frac{K}{2}h^2
        \right)}
        \sum_{m=0}^{N-1}
        e^{-\imath2\pi\frac{Kh}{N}m}.
    \end{aligned}
\end{equation}
This shows that  $|\mathcal{G}_{u,j}[r,q]|^2$ is composed of a quadratic Gauss sum term and a Dirichlet sum. The Dirichlet sum satisfies
\begin{equation}\label{equ:appendix_Dirichlet_sum}
    \sum_{m=0}^{N-1}
    e^{-\imath\frac{2\pi}{N}Khm}
    =
    \begin{cases}
    N, & N\mid Kh,\\
    0, & N\nmid Kh.
    \end{cases}
\end{equation}

Let $K_0=K/D\in \mathbb{Z}$ and $N_0=N/D\in \mathbb{N}$. Since $\gcd(|K_0|,N_0)=1$, the divisibility condition $N\mid Kh$ in \eqref{equ:appendix_Dirichlet_sum} is equivalent to $N_0\mid h$. As $h=(n-m)(\mathrm{mod}\,N)\in\{0,1,\ldots,N-1\}$, the divisibility condition $N_0\mid h$ occurs only at
\begin{equation}\label{equ:appendix_h}
    h=dN_0,\quad d=0,1,\ldots,D-1,
\end{equation}
which is also the nonzero term condition of the Dirichlet sum in \eqref{equ:appendix_Dirichlet_sum}. Substituting \eqref{equ:appendix_h} into \eqref{equ:appendix_G_square_decomposed}, we obtain
\begin{equation}\label{equ:appendix_G_A_K_further}
    \begin{aligned}
    |\mathcal{G}[A]|^2
    &=
    N\sum_{d=0}^{D-1}
    e^{-\imath 2\pi
    \left(
    \frac{A}{D}d
    +
    \frac{K_0N_0}{2}d^2
    \right)} .
    \end{aligned}
\end{equation}

Because $D<N$ and $N=2^n$, $N_0=N/D\geq 2$ is even. Hence, $({K_0N_0}/{2})d^2\in\mathbb{Z}$ for all $d$ in \eqref{equ:appendix_G_A_K_further}, so the corresponding quadratic phase factor equals one. The remaining summation is again a Dirichlet term, given as
\begin{equation}\label{equ:appendix_G_A_K_lemma1}
    |\mathcal{G}[A]|^2
    =
    N\sum_{d=0}^{D-1}
    e^{-\imath 2\pi \frac{A}{D}d}
    =
    \begin{cases}
    ND, & D\mid A\\
    0, & D\nmid A
    \end{cases}.
\end{equation}

Finally, $A=r-q$ converts the two divisibility conditions into $r\equiv q \pmod{D}$ and $r\not\equiv q \pmod{D}$, respectively. Restoring $D=D_{u,j}$ and $\mathcal{G}[A]=\mathcal{G}_{u,j}[r,q]$ in \eqref{equ:appendix_G_A_K_lemma1} yields \eqref{equ:lemma1}. This completes the proof of \textit{Lemma \ref{lemma1}}.

\section{Proof of Theorem 3}
\label{app:theorem3}

To prove \textit{Theorem~\ref{theorem3}}, we show that $\{\mathcal I_u\}_{u=1}^{U}$ satisfies the demand constraints \eqref{prob:residue_class_allo_demand}, the full-loading condition \eqref{equ:full_loading_res}, and the pairwise modular constraints \eqref{prob:residue_class_allo_conflict}.

For notational brevity, define 
$\mathcal{B}_u=\{b\in\mathcal{Q}:\phi(b)=u\}$. According to \eqref{equ:I_u}, $\mathcal I_u=\mathcal{A}_u \cup \mathcal{B}_u$.

\textbf{Demand constraints \eqref{prob:residue_class_allo_demand}}: Since $\mathcal B_u\subseteq\mathcal{Q}$ and $\mathcal Q\cap\mathcal A_u=\varnothing$, $\forall u$, we have $\mathcal A_u\cap\mathcal B_u=\varnothing$, $\forall u$. The feasible representative allocation satisfies $|\mathcal A_u|=\eta_u$, $\forall u$. Therefore,
$
|\mathcal I_u|
    =
    |\mathcal A_u|+|\mathcal B_u|
    \geq
    \eta_u,
    \forall u,
$
and \eqref{prob:residue_class_allo_demand} is satisfied.

\textbf{Full-loading condition \eqref{equ:full_loading_res}}: $\{\mathcal A_u\}_{u=1}^{U}$ are mutually disjoint, and $\mathcal Q$ contains exactly the remaining fine residue classes in $\mathbb Z_L$. Since \eqref{equ:phi_b} assigns each $b\in\mathcal Q$ to exactly one user, the final sets $\{\mathcal I_u\}_{u=1}^{U}$ form a partition of $\mathbb{Z}_L$ and satisfy the full-loading condition \eqref{equ:full_loading_res}.

\textbf{Pairwise modular constraint \eqref{prob:residue_class_allo_conflict}}: For notational brevity, fix an arbitrary user pair $u<j$ and let $D=\widetilde D_{u,j}$. Since $\mathcal I_u=\mathcal A_u\cup\mathcal B_u$ and $\mathcal I_j=\mathcal A_j\cup\mathcal B_j$, it is sufficient to verify the following four cases:
\begin{subequations}
    \begin{align}
        & g\not\equiv f\pmod{D},
        \,\,
        \forall g\in\mathcal A_u,\ 
        \forall f\in\mathcal A_j,
        \label{equ:appendix_conflict_A_A}\\
        & g\not\equiv f\pmod{D},
        \,\,
        \forall g\in\mathcal A_u,\ 
        \forall f\in\mathcal B_j,
        \label{equ:appendix_conflict_A_B}\\
        & g\not\equiv f\pmod{D},
        \,\,
        \forall g\in\mathcal B_u,\ 
        \forall f\in\mathcal A_j,
        \label{equ:appendix_conflict_B_A}\\
        & g\not\equiv f\pmod{D},
        \,\,
        \forall g\in\mathcal B_u,\ 
        \forall f\in\mathcal B_j.
        \label{equ:appendix_conflict_B_B}
    \end{align}
\end{subequations}

\textbf{Case~\eqref{equ:appendix_conflict_A_A}:} It follows from the feasibility of  $\{\mathcal A_u\}_{u=1}^{U}$.

\textbf{Case~\eqref{equ:appendix_conflict_A_B}:} Suppose, for contradiction, that there exist $g_u\in\mathcal A_u$ and $f\in\mathcal B_j$
such that $g_u\equiv f\pmod D$. For any $g_j\in\mathcal A_j$, the feasibility of the representative allocation gives
$g_u\not\equiv g_j\pmod D$. Since $f\equiv g_u\pmod D$, we have
\begin{equation}\label{equ:appendix_f_not_g_j}
    f\not\equiv g_j\pmod D,
    \quad \forall g_j\in\mathcal A_j.
\end{equation}
Since all relevant moduli in $\mathcal D$ are powers of two, any $D'\in\mathcal{D}$ with $D'\geq D$ is a multiple of $D$. Therefore, congruence modulo $D'$ would imply congruence modulo $D$. From \eqref{equ:appendix_f_not_g_j}, we have $f\not\equiv g_j\pmod {D'}$, $\forall g_j\in\mathcal A_j$, $\forall D'\in\mathcal{D}$ with $D'\geq D$. Hence, the largest modular similarity between $f$ and user $j$ must be less than $D$, i.e., 
\begin{equation}\label{equ:appendix_rho_j_f}
    \rho_j(f)=\max_{g\in\mathcal A_j}\rho(g,f)<D.
\end{equation} 

On the other hand, since $f\equiv g_u\pmod D$,
\begin{equation}
    \rho_u(f)
    =
    \max_{g\in\mathcal A_u}\rho(g,f)
    \geq
    \rho(g_u,f)
    \geq D. 
\end{equation}
Thus, $\rho_u(f)>\rho_j(f)$. According to the extension rule $\phi(f)=\operatorname{TB}\big(\arg\max_{1\leq i\leq U}\rho_i(f)\big)$, $f$ cannot be assigned to user $j$, which contradicts $f\in\mathcal{B}_j$.
Therefore, Case~\eqref{equ:appendix_conflict_A_B} holds.

\textbf{Case~\eqref{equ:appendix_conflict_B_A}:} It follows symmetrically from Case~\eqref{equ:appendix_conflict_A_B}.

\textbf{Case~\eqref{equ:appendix_conflict_B_B}:} Suppose, for contradiction, that there exist $g_u\in\mathcal B_u$ and
$f_j\in\mathcal B_j$ such that $g_u\equiv f_j\pmod D$.
By Case~\eqref{equ:appendix_conflict_A_B}, $f_j$ is not congruent modulo $D$ to any element in $\mathcal A_u$. Since $g_u\equiv f_j\pmod D$, we have
$g_u\not\equiv g\pmod D$, $\forall g\in\mathcal A_u$. Following the same argument as in \eqref{equ:appendix_f_not_g_j}--\eqref{equ:appendix_rho_j_f}, we obtain $\rho_u(g_u)<D$. Since $g_u$ is assigned to user $u$, user $u$ attains the maximum modular similarity to $g_u$, and thus
\begin{equation}
    \rho_t(g_u)\leq \rho_u(g_u)<D,
    \quad \forall t .
\end{equation}
Similarly, using Case~\eqref{equ:appendix_conflict_B_A}, we have $\rho_t(f_j)<D$, $\forall t$.

We now show that $\rho_t(g_u)=\rho_t(f_j)$ for every user
$t$. Let $D_t=\rho_t(g_u)$. Since $D_t<D$ and both are powers of two, $D_t\mid D$. From
$g_u\equiv f_j\pmod D$, we have
$g_u\equiv f_j\pmod {D_t}$. By the definition of
$D_t=\rho_t(g_u)$, there exists $a_t\in\mathcal A_t$ attaining this maximum, such that $g_u\equiv a_t\pmod {D_t}$. Therefore, $f_j\equiv a_t\pmod {D_t}$, which implies $\rho_t(f_j)\geq D_t=\rho_t(g_u)$. Reversing the roles of $g_u$ and $f_j$ gives $\rho_t(g_u)\geq \rho_t(f_j)$. Hence,
\begin{equation}
    \rho_t(g_u)=\rho_t(f_j),
    \quad \forall t .
\end{equation}
Therefore, $g_u$ and $f_j$ have the same modular-similarity values for all users. Since the tie-breaking rule $\mathrm{TB}(\cdot)$ in \eqref{equ:phi_b} is deterministic, the extension rule must assign them to the same user, i.e., $\phi(g_u)=\phi(f_j)$, which contradicts $g_u\in\mathcal B_u$, $f_j\in\mathcal B_j$, and $u\neq j$. Thus, Case~\eqref{equ:appendix_conflict_B_B} holds.

Thus, all pairwise modular constraints are satisfied, completing the proof of \textit{Theorem~\ref{theorem3}}.

\ifCLASSOPTIONcaptionsoff
  \newpage
\fi

\bibliographystyle{IEEEtran}
\bibliography{references}

\end{document}